\documentclass{nature_arxiv}
\usepackage{hyperref}
\pdfoutput=1

\title{Multi-cancer molecular signatures and their interrelationships}

\author{Wei-Yi Cheng$^{1}$, Tai-Hsien Ou Yang$^1$, Hui Shen$^2$, Peter W. Laird$^2$, Dimitris Anastassiou$^1$ and the Cancer Genome Atlas Research Network}

\begin{document}

\begin{raggedright}

\maketitle

\begin{affiliations}
 \item Columbia Initiative in Systems Biology and Department of Electrical Engineering, Columbia University, New York, NY, USA
 \item USC Epigenome Center, Norris Comprehensive Cancer Center, Keck School of Medicine, University of Southern California, Los Angeles, CA, USA.
\end{affiliations}


\vspace{2in}
\begin{addendum}
 \item[Corresponding Author] Dimitris Anastassiou, Columbia Initiative in Systems 
Biology and Department of Electrical Engineering, Columbia University, 1312 S.W. Mudd Building, 
-- Mail Code 4712, 500 West 120th Street, USA. Phone: +1‑212‑854-3113; E-mail: da8@columbia.edu
\vspace{0.2in}
\item[] \textbf{ This is version 2. A previous version of the same article appears at \url{http://arxiv.org/pdf/1306.2584v1.pdf} dated June 11, 2013. }
\item[] \textbf{ 
Please note: The pan-cancer molecular signatures disclosed in this article are the results of applying 
our data mining algorithm to the rich TCGA ``pancan12'' data sets from twelve different cancer types. 
These signatures have been identified as patterns, without any indication about their role or potential 
usefulness. We believe that each of them represents an important biomolecular event in cancer.
}
\item[] \textbf{
We invite the cancer research community to contact us and help us interpret each of these pan-cancer signatures, and to investigate 
them for potential applications in diagnostic, prognostic and therapeutic products applicable to multiple 
cancer types. There will be additional versions of this article and we expect that the final version will contain many co-authors.
}
\end{addendum}

\pagebreak
\begin{abstract}
Although cancer is known to be characterized by several unifying biological hallmarks, systems biology has had 
limited success in identifying molecular signatures present in in all types of cancer. The current availability 
of rich data sets from many different cancer types provides an opportunity for thorough computational data mining 
in search of such common patterns. Here we report the identification of 18 ``pan-cancer'' molecular signatures resulting 
from analysis of data sets containing values from mRNA expression, microRNA expression, DNA methylation, and protein 
activity, from twelve different cancer types. The membership of many of these signatures points to particular biological 
mechanisms related to cancer progression, suggesting that they represent important attributes of cancer in need of 
being elucidated for potential applications in diagnostic, prognostic and therapeutic products applicable to multiple 
cancer types.
\end{abstract}

Cancer is known to be not just one disease, but many diseases, as evidenced by the diversity of its 
pathological manifestations. On the other hand, it has been appreciated that there exist some unifying 
capabilities, or ``hallmarks,'' characterizing all cancers, as proposed in two seminal papers\cite{hanahan2000, hanahan2011}. It is 
reasonable to hypothesize that such common biological traits would be represented by particular patterns 
detectable in data sets derived from cancer samples. However, systems biology has had limited success in 
finding such common patterns until recently. The current availability of integrated biomolecular data sets 
from twelve cancer types (``pancan12'') in The Cancer Genome Atlas (TCGA) provides an opportunity for 
thorough data mining, so that such common patterns can be computationally discovered and defined with high accuracy.

Our data mining approach\cite{mePLoS} uses an iterative algorithm to identify patterns that manifest themselves as 
distinct molecular signatures, called attractor metagenes, several of which were found in nearly identical form following 
separate analysis of data sets from multiple different cancer types. The algorithm is designed to converge to the core of 
gene coexpression patterns, without being influenced in any way by other constraints, such as classification of samples into 
subtypes. These signatures are manifested by the coordinate observed presence of many features (such as expression of 
genes or methylation of genomic sites), to varying degrees, in multiple cancer types. The three main molecular signatures 
that we previously found\cite{mePLoS} using data sets from three cancer types are associated with mitotic chromosomal 
instability (CIN), mesenchymal transition (MES) and a lymphocyte-specific immune recruitment (LYM).  We hypothesized 
that these molecular signatures represent important biomolecular events of cancer, and therefore that they would be associated 
with phenotypes in multiple cancer types. Consistent with this hypothesis, a computational model using attractor metagenes 
as features recently won the Sage Bionetworks/DREAM Breast Cancer Prognosis Challenge\cite{meSTM, bib5}.

Here we report our results of discovering ``Pan-Cancer'' molecular signatures applying the same computational methodology 
(\textbf{Methods}) on the TCGA pancan12 data sets. Based on parameter choices that would guarantee that such signatures 
are clearly present in the majority of the data sets and would involve a significant number of mutually associated genes.

\section*{RESULTS}
\subsection{Listing of 18 Pan-Cancer signatures}
We identified 15 attractor molecular signatures, seven of which were 
present in mRNA expression data sets, three in DNA methylation data sets, 
three in microRNA expression data sets, and two in protein activity data 
sets.  We found several additional genomically co-localized molecular 
signatures, mainly representing amplicons, and we report on three of them,
 for a total of 18 attractor signatures. 

The signatures identified separately in individual cancer types  
available under Synapse ID syn1899444. The consensus ranked lists for 
each of these signatures are presented in \textbf{Table ~\ref{tab:tabS1}}, as well as 
under Synapse ID syn1899445. We also identified genomically co-localized 
molecular signatures, presented under synapse ID syn1899446. The first two
 such signatures (MHC Class II and GIMAP gene cluster) are strongly 
associated with the LYM metagene. The third signature contains the 
Pan-Cancer chr8q24.3 amplicon, which we had previously identified\cite{mePLoS} 
as the strongest amplicon attractor metagene.

We confirmed that the three main attractor metagenes (CIN, MES, LYM) that 
we previously identified\cite{mePLoS} are the most prominent ones (using 
a measure of signature strength defined in \textbf{Methods}) among all 18 
signatures. In addition, we identified several new attractor metagenes 
resulting from this new thorough analysis, one of which (END) contains 
endothelial markers and is associated with angiogenesis. 

\begin{figure}[!t]
\fbox{
\begin{minipage}{6.5in}
\includegraphics[width=\textwidth]{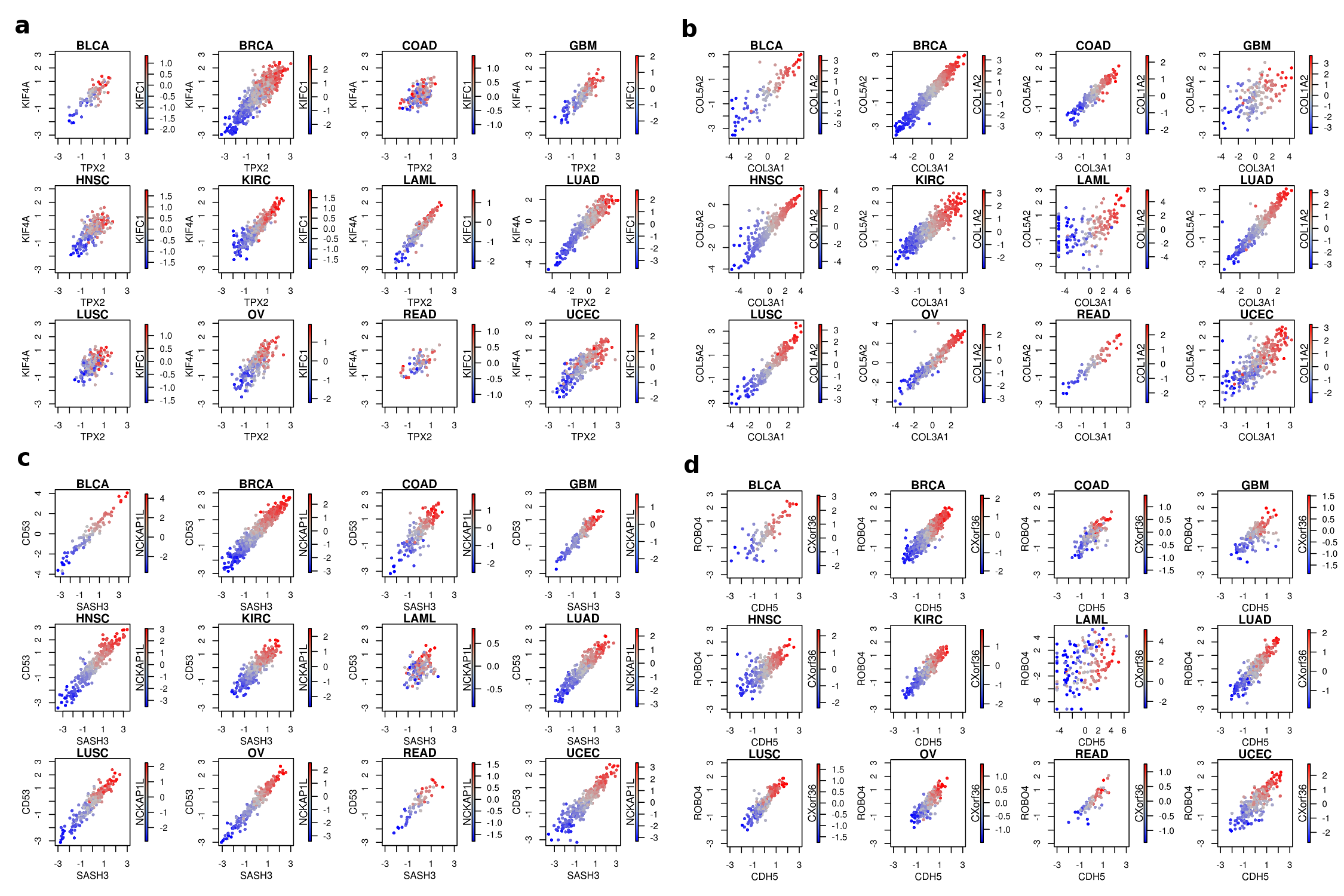}
\caption{Scatter plots of three genes from twelve cancer types. Each dot represents a cancer sample. The horizontal and vertical axes measure the expression values of two of the three genes, while the value of the third gene is color-coded. The observed linear change from lower left (blue) to upper right (red) demonstrates the coexpression of these three genes. Shown are scatter plots for the top-ranked three genes of (\textbf{a}) the CIN metagene, (\textbf{b}) the MES metagene, (\textbf{c}) the LYM metagene and (\textbf{d}) the END metagene.}
\label{fig:fig1}
\end{minipage}
}
\end{figure}

A striking visualization consistent with the co-expression of these 
Pan-Cancer molecular signatures can be made in the form of scatter plots.
 For example, \textbf{Fig. ~\ref{fig:fig1}} shows such color-coded scatter 
plots for the four main attractor metagenes CIN, MES, LYM, and END, in 
all twelve cancer types using the three top-ranked genes for each of these
 four signatures. In each scatter plot, samples represented by dots at the
 lower left (blue) side have low levels of the signature, while samples 
represented by dots at the upper right (red) side have high levels of the 
signature. \textbf{Fig. ~\ref{fig:figS1}} shows the corresponding scatter plots for all 
18 identified attractor molecular signatures demonstrating such 
coexpression in all cases. Scrutinizing each of these molecular signatures
 provides opportunities for discovery in cancer biology. 

\textbf{Table 1} provides a summary of the 18 signatures, including 
brief comments and a listing of their top-ranked members. In the following
, we briefly describe each of them.


\pagebreak
\setcounter{table}{0}
\makeatletter 
\renewcommand{\thetable}{\@arabic\c@table} 
\makeatother

\setlength{\LTcapwidth}{\textwidth}
\renewcommand{\arraystretch}{1.5}
\begin{longtable}{|p{1in}|p{4in}|p{1.5in}|}
\caption{Listing of the 18 attractor signatures}\\
\hline
\textbf{Name} & \textbf{Top members} & \textbf{Comments}\\
\hline
\endfirsthead

\multicolumn{3}{c}
{{\bfseries \tablename \thetable{} -- continued from previous page}} \\
\hline 
\textbf{Name} & \textbf{Top members} & \textbf{Comments}\\ 
\hline 
\endhead

\hline \multicolumn{3}{r}{{Continued on next page...}} \\
\endfoot

\hline \hline
\endlastfoot

\hline
\multicolumn{3}{|l|}{\textbf{mRNA}}\\
\hline
\textbf{LYM}	&	\textit{SASH3, CD53, NCKAP1L, LCP2, IL10RA, PTPRC, EVI2B, BIN2, WAS, HAVCR2}	&	lymphocyte infiltration\\
\hline
\textbf{CIN}	&	\textit{TPX2, KIF4A, KIFC1, NCAPG, BUB1, NCAPH, CDCA5, KIF2C, PLK1, CENPA}	&	mitotic chromosomal instability\\
\hline
\textbf{MES}	&	\textit{COL3A1, COL5A2, COL1A2, THBS2, COL5A1, VCAN, COL6A3, SPARC, AEBP1, FBN1}	&	mesenchymal transition \\
\hline
\textbf{END}	&	\textit{CDH5, ROBO4, CXorf36, CD34, CLEC14A, ARHGEF15, CD93, LDB2, ELTD1, MYCT1} 	&	endothelial markers\\
\hline
\textbf{``\textit{AHSA2}''}	&	\textit{AHSA2, LOC91316, PILRB, ZNF767, TTLL3, CCNL2, PABPC1L, LENG8, CHKB CPT1B, SEC31B}	&\\
\hline
\textbf{IFIT}	&	\textit{IFIT3, MX1, OAS2, RSAD2, CMPK2, IFIT1, IFI44L, IFI44, IFI6, OAS1}	&	interferon-induced\\
\hline
\textbf{``\textit{WDR38}''}	&	\textit{WDR38, YSK4, ROPN1L, C1orf194, MORN5, WDR16, RSPH4A, FAM183A, ZMYND10, DNAI1}	&	\\
\hline
\multicolumn{3}{|l|}{\textbf{Genomically co-localized mRNA}} \\
\hline
\textbf{MHC Class II}	&	\textit{HLA-DPA1, HLA-DRA, HLA-DPB1, HLA-DRB1, HLA-DMA, HLA-DMB, HLA-DOA, HLA-DQA1, HLA-DRB5}	&	strongly associated with LYM\\
\hline
\textbf{GIMAP cluster}	&	\textit{GIMAP4, GIMAP7, GIMAP6, GIMAP5, GIMAP8, GIMAP1}	&	strongly associated with LYM\\
\hline
\textbf{Chr8q24.3 amplicon}	&	\textit{SHARPIN, HSF1, TIGD5, GPR172A, ZC3H3, EXOSC4, SCRIB, CYHR1, MAF1, PUF60} 	&	most prominent Pan-Cancer amplicon\\
\hline
\multicolumn{3}{|l|}{\textbf{microRNA}} \\
\hline
\textbf{\textit{DLK1}-\textit{DIO3} RNA cluster}	&	mir-127, mir-134, mir-379, mir-409, mir-382, mir-758, mir-381, mir-370, mir-654, mir-431	&	includes \textit{MEG3} long noncoding RNA; associated with MES\\
\hline
\textbf{``mir-509''}	&	mir-509, mir-514, mir-508	&	\\
\hline
\textbf{``mir-144''}	&	mir-144, mir-451, mir-486	&	associated with erythropoiesis\\
\hline
\multicolumn{3}{|l|}{\textbf{Methylation}}\\
\hline
\textbf{``RMND1''}	&	RMND1-6-151814639, MAP3K7-6-91353911, DNAAF1-16-82735714, PTRH2-17-55139429, ZNF143-11-9439170, cg03627896
, TAMM41-3-11863582, CDK5-7-150385869, OTUB1-11-63510174, AATF-17-32380976	&	\\
\hline
\textbf{M+}	&	cg13928306, MTMR11-1-148175405, cg27324619, TNKS1BP1-11-56846646, C11orf52-11-111294703, IL17RC-3-9934128, cg24765079, ERBB3-12-54759072, IL22RA1-1-24342151, C11orf52-11-111294903	&	methylated in infiltrating lymphocytes\\
\hline
\textbf{M-}	&	BIN2-12-50003941, PTPRCAP-11-66961771, TNFAIP8L2-1-149395922, IGFLR1-19-40925164, FAM113B-12-45896487, CD6-11-60495754, KLHL6-3-184755939, PTPN7-1-200396189, FAM78A-9-133141340, ACAP1-17-7180947	&	Unmethylated in infiltrating lymphocytes, may be causal to the expression of some of the genes of the LYM signature\\
\hline
\multicolumn{3}{|l|}{\textbf{Protein activity}} \\
\hline
\textbf{``c-Met''}	&	c-Met, Snail, PARP\_cleaved, Caspase-8, ERCC1, Rb	&	Related to apoptosis\\
\hline
\textbf{``Akt''}	&	Akt, Tuberin, STAT5A	&\\
\end{longtable}

\begin{figure}[!p]
\fbox{
\begin{minipage}{6.5in}
\includegraphics[width=\textwidth]{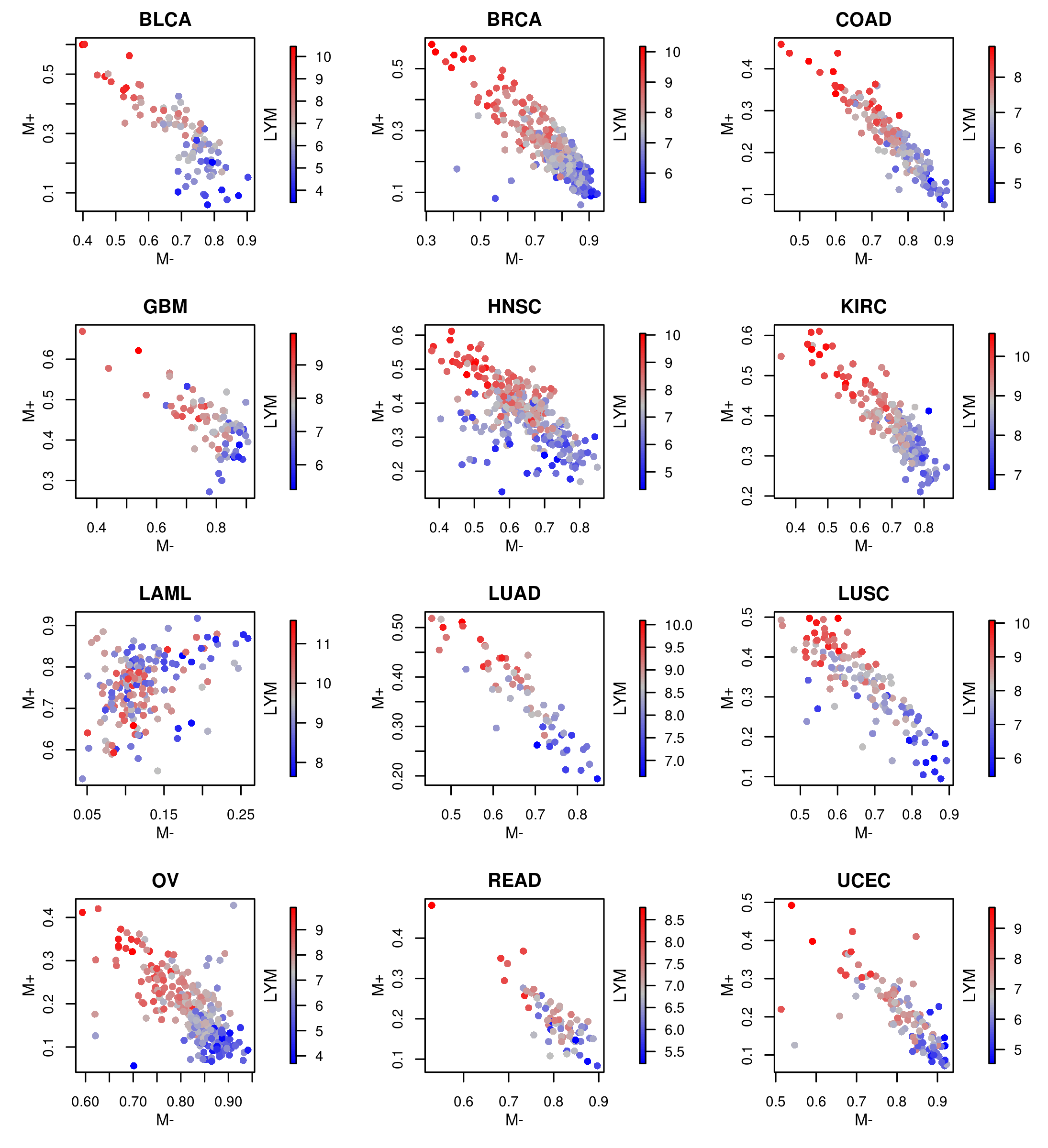}
\caption{
Scatter plots connecting the LYM, M+ and M- metagenes in 12 cancer types. Each dot represents a cancer sample. The horizontal and vertical axes measure the average methylation values of the two methylation signatures, M- and M+, while the value of the expression of the LYM metagene is color-coded. In all three cases, the metagene is defined by the average of the top ranked genes as described in \textbf{Table ~\ref{tab:tabS1}}. 
}
\label{fig:fig2}
\end{minipage}
}
\end{figure}

\subsection{Lymphocyte infiltration: LYM mRNA signature; M+\hspace{1pt} methylation signature; M-\hspace{1pt} methylation signature}
These three signatures are related to tumor infiltration by lymphocytes. We list 
them together because they are strongly interrelated (\textbf{Fig. ~\ref{fig:fig2}}) even though each 
of the three was independently derived using an unsupervised computational method. 
The presence of LYM is accompanied by the presence of M+ and the absence of M- in 
all solid cancer types, suggesting that the three signatures reflect the same 
biomolecular event, which appears to be the infiltration of immune cells in tumor 
tissue. Indeed, there is remarkable similarity (\textbf{Fig. ~\ref{fig:fig3}}) between the LYM signature 
and the ``immune score'' of the ESTIMATE tumor purity computational tool (\url{http://ibl.mdanderson.org/estimate}). 
The values of the M+ methylation signature are also remarkably similar to those of the methylation-based 
``leukocyte percentage'' estimation \cite{huiNature} (available under Synapse ID syn1809222).

\begin{figure}[!p]
\fbox{
\begin{minipage}{6.5in}
\includegraphics[width=\textwidth]{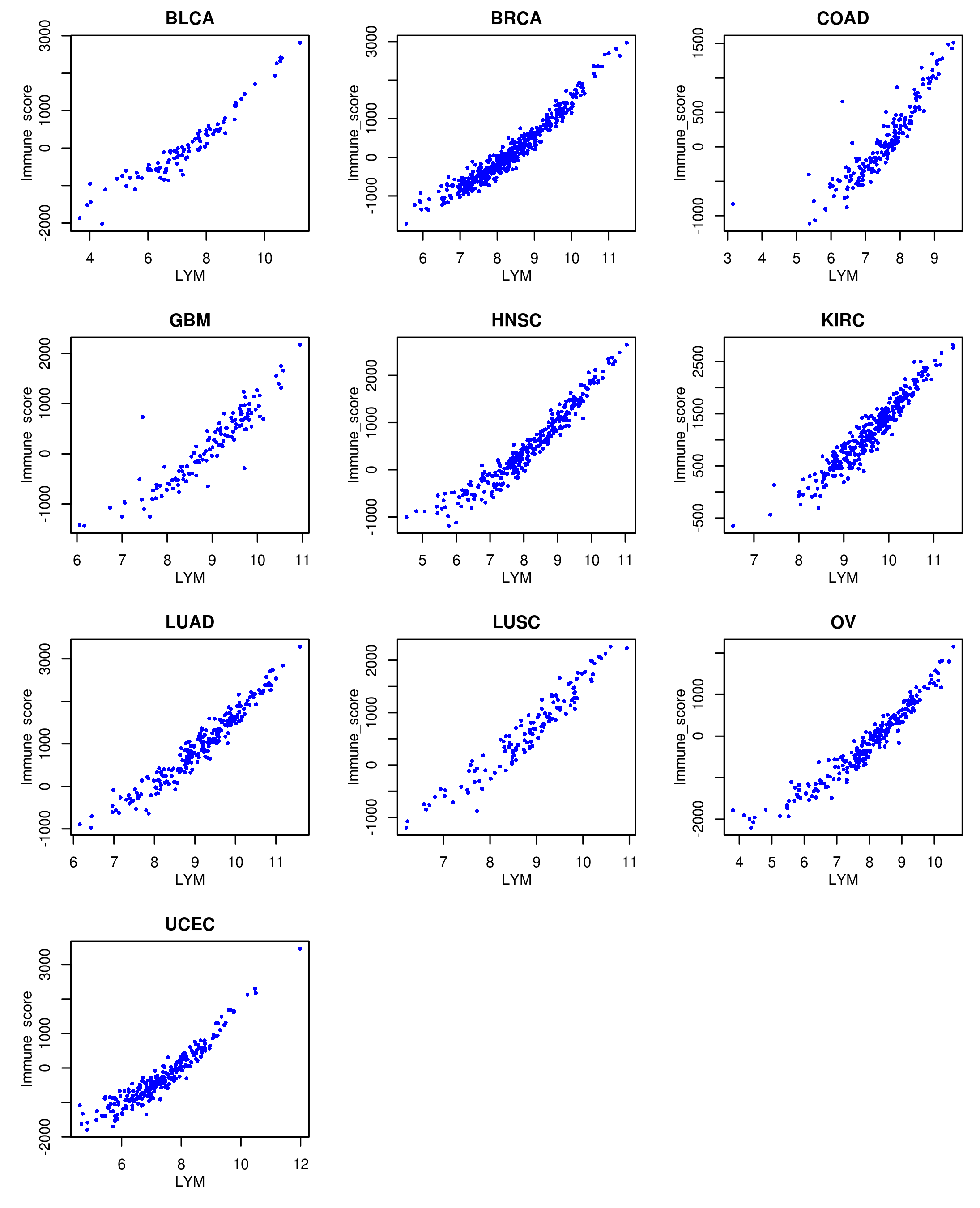}
\caption{
Scatter plots demonstrating the pan-cancer similarity of the value of the LYM metagene with the immune score of the ESTIMATE tumor purity computational tool (http://ibl.mdanderson.org/estimate) measuring immune cell infiltration. Each dot represents a cancer sample. The horizontal axis measures the expression value of the LYM metagene and the vertical axis measures the ESTIMATE immune score of infiltration. Note that the ESTIMATE did not provide scores for rectum cancer, and the estimation of immune cell infiltration is not applicable in leukemia.
}
\label{fig:fig3}
\end{minipage}
}
\end{figure}

We had previously found\cite{billCancerInfo} all three LYM, M+ and M- signatures from their association with 
the expression of miR-142. We have now confirmed this association with miR-142 in the pancan12 data sets, 
and we found that miR-150 and miR-155 are also strongly associated with the LYM signature. 
We had also previously independently identified the LYM signature as an attractor metagene\cite{mePLoS}, and 
used it in the winning model of the Sage Bionetworks Breast Cancer Prognosis Challenge\cite{meSTM}. 
Specifically the LYM signature is strongly associated with improved prognosis in ER-negative breast cancers, 
and this fact also provides an explanation for the relatively better prognosis in medullary, 
compared with other types of high-grade breast cancers. 

The interrelationship of the LYM, M+ and M- signatures, as shown in \textbf{Fig. ~\ref{fig:fig2}}, appears to be a consequence 
of the presence of different subclasses of cells (as opposed to being a methylation switch inside 
the same cell), consistent with their assumed role of measuring the extent of lymphocyte infiltration 
in the tumor. In other words, the M+ methylation sites, normally unmethylated, are largely methylated 
in the infiltrating leukocytes; and the M- methylation sites, normally methylated, are largely 
unmethylated in the infiltrating leukocytes. Consistently, many of the genes methylated by the M- 
signature are identical to those of LYM (six among the 27 genes of the M- signature (\textit{BIN2}, \textit{TNFAIP8L2}, 
\textit{ACAP1}, \textit{NCKAP1L}, \textit{FAM78A}, \textit{PTPN7}) listed in Table S1 are also among the 168 
genes listed in the LYM attractor metagene 
($P < 9.21\times10^{-7}$ based on Fisher’s exact test). The observed significant overlap in the gene sets and 
the negative association between gene expression in LYM and DNA methylation in M- are consistent with the notion 
that the absence of DNA methylation is permissive for gene expression, suggesting that the expression of the LYM
signature in the infiltrating lymphocytes may be facilitated in part by the hypomethylation of the M- signature.

The sharp definition of the LYM signature (being a Pan-Cancer attractor signature pointing to few genes 
at the core of coexpression) provides strong hints about the precise nature of this leukocyte infiltration. 
Specifically, the membership of the top-ranked genes (\textit{SASH3}, \textit{CD53}, \textit{NCKAP1L}, \textit{LCP2}, 
\textit{IL10RA}, \textit{PTPRC}, \textit{EVI2B}, \textit{BIN2}, \textit{WAS}, \textit{HAVCR2}, \ldots) 
point to a specific type of lymphocytes. We have speculated\cite{mePLoS} that these infiltrating lymphocytes 
are T cells having undergone a particular type of co-stimulation providing hypotheses for related adoptive transfer therapy.

Two proteins strongly associated with the LYM signature are two tyrosine kinases: Lck (lymphocyte-specific protein tyrosine kinase) 
and Syk (spleen tyrosine kinase).

\subsection{CIN (mitotic chromosomal instability) mRNA signature}
This signature is related to mitotic chromosomal instability. It is similar to numerous known ``proliferation'' signatures, 
but its sharp definition as an attractor metagene specifically points to the kinetochore-microtubule interface and 
associated kinesins. Comparison with similar mitotic signatures in normal cells may help pinpoint driver genes for 
malignant chromosomal instability. The signature is strongly associated with tumor grade as well as poor prognosis in many, 
if not all, cancer types.

Two proteins strongly associated with the CIN signature are Cyclin B1 and CDK1. Consistently, it is known that the cyclin 
B1-Cdk1 complex of cyclin-dependent kinase 1 is involved in the early events of mitosis, and that nuclear cyclin B1 protein 
may induce chromosomal instability and enhance the aggressiveness of the carcinoma cells\cite{bibi7}. 

\subsection{MES (mesenchymal transition) mRNA signature}
This signature is related to mesenchymal transition and invasiveness of cancer cells. It is similar to numerous 
``stromal'' or ``mesenchymal'' signatures; however there is evidence\cite{dimitrisBMCCancer} that many among the genes of the 
signature are largely produced by transdifferentiated cancer cells. We hypothesize that such cells, known to 
assume the duties of cancer-associated fibroblasts in some tumors\cite{hanahan2011}, may  have become indistinguishable, 
even using laser capture microdissection, from stromal fibroblasts. We had originally identified the MES signature 
from its association with tumor stage\cite{hoonBMCMedGenomics}; specifically the signature appears only after a particular cancer 
type-specific tumor stage threshold has been reached.

The values of the MES signature are remarkably similar to the ``stromal score'' of the ESTIMATE tumor purity computational tool 
(\url{http://ibl.mdanderson.org/estimate}) measuring fibroblast infiltration. Based on our previous reasoning, however, 
we believe that this interpretation may not be fully accurate, and that it will be important to find out to what extent 
some of the cells expressing some of these mesenchymal markers may actually be transdifferentiated cancer cells, and 
whether the estimated tumor purity may be affected by other types of normal cells instead of stromal fibroblasts.

The co-regulated microRNAs most strongly associated with the MES signature are miR-199a, miR-199b, and miR-214. The \textit{DLK1}-
\textit{DIO3} RNA cluster attractor signature, described later, is also strongly associated with MES.

The protein most strongly associated with the MES signature is Fibronectin.

\subsection{END (endothelial marker) mRNA signature}
This is a novel angiogenesis-associated attractor signature. Nearly all the top-ranked genes (\textbf{Table 1}) are 
endothelial markers.  The top gene, \textit{CDH5}, codes for VE-cadherin, which is known to be involved in a pathway suppressing 
angiogenic sprouting\cite{bib9}. The second gene, \textit{ROBO4}, is known to inhibit VEGF-induced pathologic angiogenesis and endothelial 
hyperpermeability\cite{bib10}. Consistently, the END attractor metagene appears to be protective and anti-angiogenic, stabilizing the 
vascular network. For example, 22 out of the 27 genes of the END attractor are among the 265 genes included in File S2 of a recent 
study\cite{bib11} of renal cell carcinoma ($P < 8.4\times10^{-38}$ based on Fisher’s exact test) as most associated with patients’ 
survival. These good-prognosis genes were intermixed in the same file with many poor-prognosis genes of the CIN attractor, 
suggesting that the CIN and END attractor metagenes are two of the most prognostic features in renal cell carcinoma. 

Interestingly, the MES and END attractor metagenes are positively associated with each other (\textbf{Fig. ~\ref{fig:fig4}}), in the sense 
that overexpression of the END signature tends to imply overexpression of the MES signature and vice-versa. This is consistent 
with mutual exclusivity between angiogenesis and invasiveness and with related findings\cite{bib12} that VEGF inhibits tumor cell 
invasion and mesenchymal transition, while antiangiogenic therapy is associated with increased invasiveness\cite{bib13}. It may also 
explain the paradoxical protective nature of signatures related to the MES attractor metagene in invasive breast cancers\cite{bib14}.

\begin{figure}[!p]
\fbox{
\begin{minipage}{6.5in}
\includegraphics[width=\textwidth]{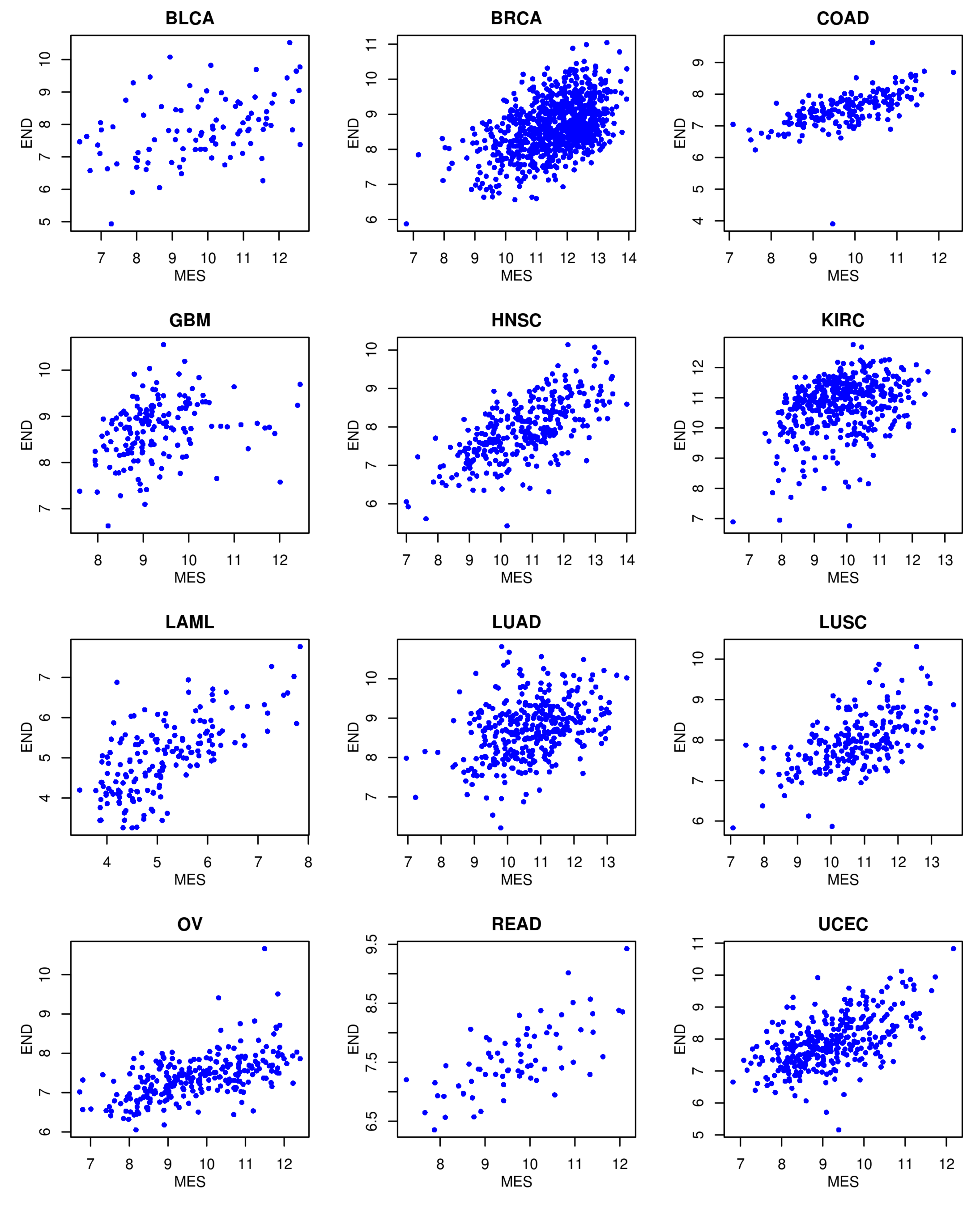}
\caption{
Scatter plots demonstrating the association between MES and END attractor metagenes. The horizontal and vertical axes measure the values of the MES and END signatures. The two signatures have positive correlation, although this association is not sufficiently strong to merge the two attractors into one. This association suggests that the invasive MES signature and the antiangiogenic END signature tend to be present simultaneously.
}
\label{fig:fig4}
\end{minipage}
}
\end{figure}

\subsection{``AHSA2'' mRNA signature}
We do not yet know what this signature represents. We observed that several noncoding RNAs (e.g. NCRNA00105, NCRNA00201) 
are in relatively high-ranked positions among its members.

\subsection{IFIT (interferon-induced) mRNA signature}
The members of this signature are interferon-induced. For example, we observed large enrichment of the genes of the 
signature among those upregulated by IFN-$\alpha$ in the side population (SP) of ovarian cancer cells\cite{bib15} from the 
list provided in Supplementary Table S4 of that paper, in which the authors concluded that tumors bearing large SP numbers 
could be particularly sensitive to IFN-$\alpha$ treatment. 

\subsection{``WDR38'' mRNA signature}
We do not know what this signature represents, except that we had found one of its key members, gene \textit{ZMYND10}, to be 
protective and associated with estrogen receptor expression in breast cancer.

\subsection{MHC Class II genomically co-localized mRNA signature}
We found this signature using the genomically co-localized version of the algorithm. It is very highly correlated with LYM.

\subsection{GIMAP genomically co-localized mRNA signature}
As above, we found this signature using the genomically co-localized version of the algorithm. It is also very highly correlated with LYM.

\subsection{Chr8q24.3 amplicon mRNA signature}
This is the strongest pan-cancer amplicon signature. It was previously found predictive of early relapse in ER-positive breast cancers\cite{bib16}.

\subsection{``RMND1'' methylation signature}
We do not yet know what the comethylation of the sites of this signature signifies.

\begin{figure}[!t]
\fbox{
\begin{minipage}{6.5in}
\includegraphics[width=\textwidth]{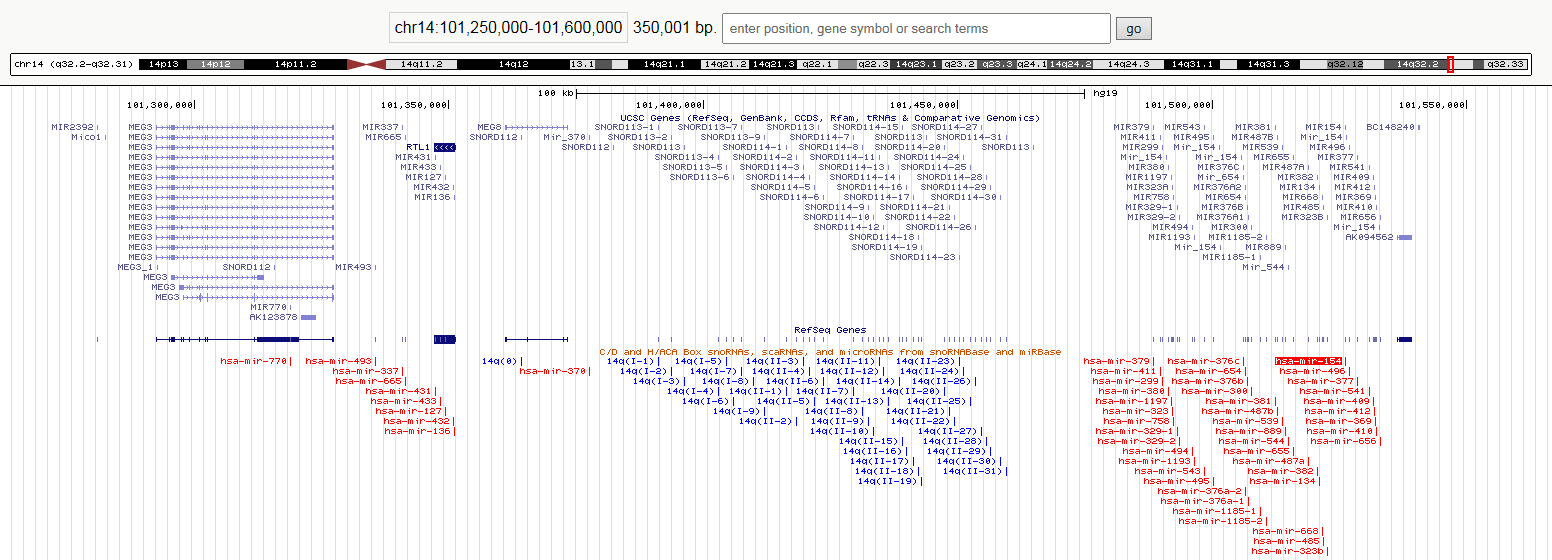}
\caption{
The \textit{DLK1}-\textit{DIO3} cluster of noncoding RNAs. Shown is a screen capture from the UCSC Genome Browser (\url{http://genome.ucsc.edu}). The cluster of imprinted genes delineated by the \textit{DLK1} and \textit{DIO3} genes (outside the shown region) is located on chromosome 14. We found that the corresponding pan-cancer attractor signature does not contain any paternally inherited protein-coding genes. It does contain the numerous noncoding RNA genes expressed from the maternally inherited homolog, including the \textit{MEG3} long noncoding RNA gene.
}
\label{fig:fig5}
\end{minipage}
}
\end{figure}

\subsection{\textit{DLK1}-\textit{DIO3} RNA cluster signature}
This is the strongest pan-cancer multi-microRNA coexpression signature. It consists of numerous noncoding RNAs within the 
\textit{DLK1}-\textit{DIO3} imprinted genomic region of chr14q32. \textbf{Fig. ~\ref{fig:fig5}} shows a screen capture of the genomic region 
from the UCSC Genome Browser (\url{http://genome.ucsc.edu/}). We confirmed that the coexpression signature also includes the 
\textit{MEG3} long noncoding RNA located at the upstream end of the region. It may also include numerous small nuclear 
RNAs at the central region, but there were no associated probe sets to confirm the coexpression. We found that this ncRNA 
signature is associated with the MES (mesenchymal transition) mRNA signature. For example, the ranked list of mRNAs most 
associated with the \textit{DLK1}-\textit{DIO3} ncRNA signature starts from \textit{POSTN}, \textit{PCOLCE}, \textit{COL5A2}, \textit{COL1A2}, 
\textit{GLT8D2}, \textit{COL5A1}, \textit{SFRP2}, and \textit{FAP}.

Expression of the imprinted \textit{DLK1}-\textit{DIO3} ncRNA cluster is believed to be vital for the development potential of 
embryonic stem cells\cite{bib17}, consistent with the hypothesis\cite{bib18} that mesenchymal transition in cancer reactivates 
embryonic developmental programs and makes cancer cells invasive and stem-like. The \textit{DLK1}-\textit{DIO3} ncRNA signature 
was also found to define a stem-like subtype of hepatocellular carcinoma associated with poor survival\cite{bib19}. The details 
of the regulation mechanism for this ncRNA cluster coexpression in the \textit{DLK1}-\textit{DIO3} region are unclear.

\subsection{``miR-509/miR-514/miR-508'' microRNA signature}
These three microRNAs are co-localized at chrXq27.3. We do not know what this signature signifies.

\subsection{``miR-144/miR-451/miR-486'' microRNA signature}
This is a three-microRNA signature related to erythropoiesis. The first two genes are located in the bicistronic microRNA 
locus miR-144/451, highly expressed during erythrocyte development\cite{bib20}. The mRNAs most associated to this microRNA 
signature are hemoglobin-related: \textit{HBB}, \textit{HBA1}, \textit{HBA2} and \textit{ALAS2}. The protein most associated 
with this signature is HER3. These three microRNAs were identified as promising biomarkers for detection of esophageal cancer.

\subsection{c-Met/Snail/PARP\_cleaved/Caspase-8/ERCC1/Rb protein activity signature}
This protein coexpression signature appears to combine the contribution of several pathways and we hope that a plausible 
and useful biological ``story'' will be developed based on the simultaneous activity of all these six proteins in some 
cancer samples. We note that each of these proteins\cite{bib21, bib22, bib23, bib24, bib25, bib26} has been related in various ways with resistance to chemotherapy 
or apoptosis.

\subsection{Akt/Tuberin/STAT5A protein activity signature}
We do not know what the coexpression of Akt, Tuberin, STAT5A proteins represents in cancer. It is known, however, that 
low levels of STAT5A protein in breast cancer are associated with tumor progression and unfavorable clinical outcomes\cite{bib27}.

\section*{DISCUSSION}
The Pan-Cancer nature (\textbf{Fig. ~\ref{fig:figS1}}) of each of the signatures described in this paper suggests that they represent 
important biomolecular events.  A reasonable concern is whether some of these ``pan-cancer'' signatures may instead reflect 
fundamental normal ``pan-tissue'' biological mechanisms. Even if this is true for some of these signatures, this does not 
exclude the possibility that they are aberrant and play important roles in some cancer samples.  Furthermore, this provides 
the opportunity to compare similar signatures in normal vs. malignant tissues to pinpoint potential cancer-specific genes. 

Because of its exhaustive search starting from all potential ``seeds'' in all data sets from twelve different cancer types, 
our iterative data mining algorithm is guaranteed to have identified all pan-cancer molecular signatures involving simultaneous 
presence of a large number of coordinately expressed genes, proteins, or comethylated sites. We hope that these signatures are 
further scrutinized by the medical research community for the purpose of developing potential diagnostic, predictive, and 
eventually therapeutic products applicable in multiple cancers.  

\section*{ACKNOWLEDGEMENTS}
We are thankful to Hanina Hibshoosh, Chris Miller and Gordon Mills for helpful discussions, 
which contributed to improved interpretation of the signatures disclosed in this paper.

\section*{Accessibility} All figures in this paper, including supplementary figures and tables, as well as the files of generated attractor molecular signatures, are available in Synapse under ID syn1686966. 

\section*{Data description and availability} The data sets of TCGA pancan12 freeze 4.7 used to derive the results of this paper are described and are available under Synapse ID syn300013 with doi:10.7303/syn300013. The twelve cancer types are bladder urothelial carcinoma (BLCA), breast invasive carcinoma (BRCA), colon adenocarcinoma (COAD), glioblastoma multiforme (GBM), head and neck squamous cell carcinoma (HNSC), kidney renal clear cell carcinoma (KIRC), acute myeloid leukemia (LAML), lung adenocarcinoma (LUAD) , lung squamous cell carcinoma (LUSC), ovarian serous cystadenocarcinoma (OV), rectum adenocarcinoma (READ), and uterine corpus endometrioid carcinoma (UCEC).

\pagebreak

\pagebreak
\begin{methods}
\subsection{Data normalization}
The data platform for each cancer types and its corresponding Synapse ID is given below. 

\begin{table}[!hf]
\renewcommand{\arraystretch}{1}
\hspace*{-0.5in}
\begin{threeparttable}
\begin{tabular}{ |l|c|c|c|c|}
\hline
\textbf{Molecular profile} & mRNA & Protein & miRNA & DNA methylation \\ 
\hline
\multirow{3}{*}{\textbf{Platform}} &                      & Reverse phase protein    &                 & Infinium \\
                                   &    Illumina HiSeq    & lysate microarray        & Illumina HiSeq  & HumanMethylation27 \\
                                   &                      & (RPPA)                   &                 & BeadChip \\
\hline
\textbf{Cancer type} & \multicolumn{4}{|c|}{\textbf{Synapse ID}} \\
\hline
BLCA&syn1571504&syn1681048&syn1571494&syn1889358\tnote{*}\\ 
BRCA&syn417812&syn1571267&syn395575&syn411485\\ 
COAD&syn1446197&syn416772&syn464211&syn411993\\ 
GBM&syn1446214&syn416777&NA&syn412284\\ 
HNSC&syn1571420&syn1571409&syn1571411&syn1889356\tnote{*}\\ 
KIRC&syn417925&syn416783&syn395617&syn412701\\ 
LAML&syn1681084&NA&syn1571533&syn1571536\\ 
LUAD&syn1571468&syn1571446&syn1571453&syn1571458\\ 
LUSC&syn418033&syn1367036&syn395691&syn415758\\ 
OV&syn1446264&syn416789&syn1356544&syn415945\\ 
READ&syn1446276&syn416795&syn464222&syn416194\\ 
UCEC&syn1446289&syn416800&syn395720&syn416204\\ 
\hline
\end{tabular}
\begin{tablenotes}
  \item[*] The data sets were extracted from HumanMethylation450 BeadChip.
\end{tablenotes}
\end{threeparttable}
\end{table}

For each RNA sequencing and miRNA sequencing data set, the mRNAs or miRNAs in which more than 50\% of the samples 
have zero counts were removed from the data set. All the zero counts and missing values in the data sets were 
imputed using the k-nearest neighbors algorithm as implemented in the \textit{impute} package in Bioconductor. The log2 
transformed counts were then normalized using the quantile normalization methods implemented in Bioconductor's \textit{limma} 
package. The missing values in the protein and DNA methylation data sets were also imputed using the k-nearest neighbors 
algorithm in the \textit{impute} package. We summarized the miRNA expression values by taking the average expression values of the 
miRNAs with the same gene family names. For bladder and head and neck methylation data sets, for which only the Humanmethylation450 
platform were provided, we extracted the 23,380 overlapping probes between the Humanmethylation27 and HumanMethylation450 platforms 
as new data sets for analysis.

\subsection{Finding attractors}
The iterative algorithm for finding converged attractors was previously described\cite{mePLoS} and is available as 
an R package under Synapse ID syn1123167. We used the same parameters as in our previous work. Specifically, 
we selected the value of the exponent $a$ to be 5 for mRNA sequencing, and we used the same value for miRNA 
sequencing and for DNA methylation. For protein data sets due to their smaller dimension, the exponent was set to 2. 
For genomically co-localized mRNA attractors, the parameters were set as previously defined \cite{mePLoS}.
The strength of an attractor (to be used for attractor ranking as described below) was defined as the k-th highest 
mutual information among all genes with the converged attractor. For mRNA and methylation attractors, we set $k = 10$, 
and for miRNA and protein attractors, we defined $k = 3$, because we observed that these attractors tend to consist of a smaller number of mutually associated elements.

\subsection{Clustering attractors of different cancer types}
After obtaining the converged attractors in each data set, we performed a clustering algorithm to identify extremely 
similar attractors across different cancer types, using the same algorithm as in our previous work\cite{mePLoS}. We used the top features 
– mRNAs, miRNAs, proteins, or methylation probes – in each attractor as a feature set, then performed hierarchical clustering 
on the feature sets across the cancer types, using the number of overlapping features as the similarity measure. 
The number of top features used to represent the attractor was chosen according to the distribution of the features' weights in the attractors. 
For the mRNA attractors, we used the top 20 features to create such feature sets. For the methylation attractors, 
we used top 50 features for clustering. For the miRNA and protein attractors, we used the top five features for clustering. 
We removed a methylation attractor cluster containing sites exclusively on the X or Y chromosome, because we found that their values were gender-based. 
If an attractor cluster did not contain any gene that found in at least six cancer types, it was removed from consideration.

\subsection{Creating consensus molecular signatures}
To account for the fact that some of the twelve data sets may not contain sufficient heterogeneous samples for showing 
each Pan-Cancer biomolecular event, the decision of selecting a signature was based on its clear presence in at least 
half of the cancer types, \textit{i.e.}, six different cancer types. We thus created a consensus molecular signature 
from each attractor cluster as follows: We first identified, for each cluster, six significant attractors by calculating 
the sum of the similarity measures (as defined above) between each attractor and all the other attractors, ranking the 
attractors using this quantity, and selecting the six top-ranked attractors. If an attractor cluster contained less 
than six attractors, it was removed from consideration. We then calculated the average score for each feature across 
the six attractors and ranked the features accordingly as the consensus ranking. The ranking of the features is provided in \textbf{Table ~\ref{tab:tabS1}}.

\subsection{Data visualization}
To create scatter plots for the top three features in the attractor, we median-centered the values of the features 
on both axes, so the median value for each feature in each data set is zero on the scatter plots. For the color-coded feature, 
we set the median to be gray, the minimum value to be blue, and the maximum value to be red, and interpolated the colors 
for intermediate values. For mRNA sequencing and miRNA sequencing data, the outlier values were removed, where the 
outliers were identified using the \textit{boxplot} function in R. 

\subsection{Ranking attractor clusters}
The strength of an attractor cluster was defined as the average strength of the six selected attractors in the cluster, 
as identified in the previous section. 
\end{methods}

\pagebreak
\begin{suppfigure}
\setcounter{figure}{0}
\makeatletter 
\renewcommand{\thefigure}{S\@arabic\c@figure} 
\makeatother

\begin{figure}[!hp]
\caption {Scatter plots of the top three features for each of the 15 molecular signatures, demonstrating strong mutual association in nearly all cases, with very few exceptions, usually in leukemia. Each dot represents a cancer sample. The horizontal and vertical axes measure the values of two of the three features, while the value of the third feature is color-coded from blue to red.}
\pagebreak
\label{fig:figS1}
\end{figure}
\includepdf[pages=1-18]{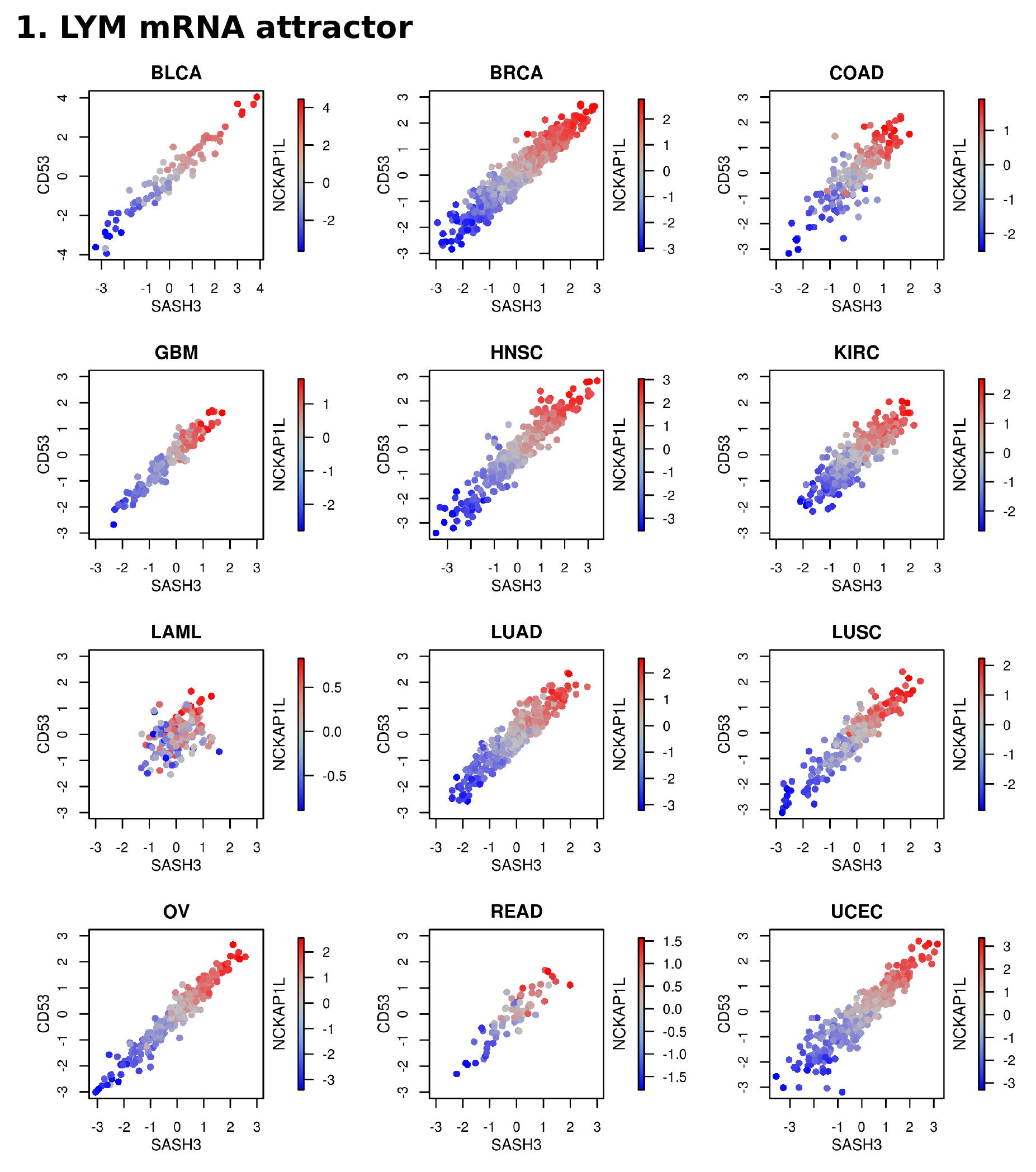}

\end{suppfigure}

\pagebreak
\begin{supptable}
\setcounter{table}{0}
\makeatletter 
\renewcommand{\thetable}{S\@arabic\c@table} 
\makeatother

%
%

\setlength{\LTcapwidth}{\textwidth}
\renewcommand{\arraystretch}{1}
\begin{longtable}{|l|c|c|l|c|c|}
\caption{The consensus rankings of features in each attractor.}\\
\hline 
\textbf{Rank} & \textbf{Gene Symbol|Entrez ID} & \textbf{Score} & \textbf{Rank} & \textbf{Gene Symbol|Entrez ID} & \textbf{Score}\\ 
\hline 
\endfirsthead

\multicolumn{6}{c}%
{{\bfseries \tablename \thetable{} -- continued from previous page}} \\
\hline 
\textbf{Rank} & \textbf{Gene Symbol|Entrez ID} & \textbf{Score} & \textbf{Rank} & \textbf{Gene Symbol|Entrez ID} & \textbf{Score}\\ 
\hline 
\endhead

\hline \multicolumn{6}{r}{{Continued on next page...}} \\
\endfoot

\hline \hline
\endlastfoot

\hline
\multicolumn{6}{|l|}{\textbf{LYM mRNA attractor}}\\
\hline
1 & SASH3|54440 & 0.848 & 85 & P2RY10|27334 & 0.582\\ 
2 & CD53|963 & 0.837 & 86 & CD300LF|146722 & 0.58\\ 
3 & NCKAP1L|3071 & 0.783 & 87 & HLA-DRA|3122 & 0.578\\ 
4 & LCP2|3937 & 0.781 & 88 & NCF1|653361 & 0.576\\ 
5 & IL10RA|3587 & 0.78 & 89 & LILRB4|11006 & 0.575\\ 
6 & PTPRC|5788 & 0.774 & 90 & FCGR1A|2209 & 0.573\\ 
7 & EVI2B|2124 & 0.765 & 91 & P2RY13|53829 & 0.571\\ 
8 & BIN2|51411 & 0.752 & 92 & SLC7A7|9056 & 0.571\\ 
9 & WAS|7454 & 0.738 & 93 & C1orf162|128346 & 0.571\\ 
10 & HAVCR2|84868 & 0.735 & 94 & C17orf87|388325 & 0.567\\ 
11 & MYO1F|4542 & 0.735 & 95 & CXorf21|80231 & 0.567\\ 
12 & CCR5|1234 & 0.735 & 96 & CYTIP|9595 & 0.567\\ 
13 & SPI1|6688 & 0.729 & 97 & NFAM1|150372 & 0.566\\ 
14 & SELPLG|6404 & 0.728 & 98 & CORO1A|11151 & 0.566\\ 
15 & CYTH4|27128 & 0.724 & 99 & GIMAP6|474344 & 0.566\\ 
16 & SLA|6503 & 0.723 & 100 & LST1|7940 & 0.566\\ 
17 & LAIR1|3903 & 0.72 & 101 & ARHGAP30|257106 & 0.564\\ 
18 & LAPTM5|7805 & 0.719 & 102 & RCSD1|92241 & 0.563\\ 
19 & PLEK|5341 & 0.712 & 103 & IL2RG|3561 & 0.562\\ 
20 & BTK|695 & 0.707 & 104 & PTPN7|5778 & 0.556\\ 
21 & FERMT3|83706 & 0.702 & 105 & FPR3|2359 & 0.555\\ 
22 & CYBB|1536 & 0.699 & 106 & CD14|929 & 0.553\\ 
23 & ITGAL|3683 & 0.698 & 107 & FYB|2533 & 0.552\\ 
24 & CD4|920 & 0.693 & 108 & GIMAP1|170575 & 0.552\\ 
25 & ARHGAP9|64333 & 0.691 & 109 & IL2RB|3560 & 0.551\\ 
26 & LILRB1|10859 & 0.688 & 110 & SH2D1A|4068 & 0.551\\ 
27 & SLAMF8|56833 & 0.685 & 111 & TNFSF13B|10673 & 0.55\\ 
28 & MPEG1|219972 & 0.683 & 112 & TRAF3IP3|80342 & 0.549\\ 
29 & C1QA|712 & 0.681 & 113 & CMKLR1|1240 & 0.546\\ 
30 & CD37|951 & 0.679 & 114 & CXCR6|10663 & 0.545\\ 
31 & ABI3|51225 & 0.679 & 115 & CCR2|729230 & 0.545\\ 
32 & MS4A6A|64231 & 0.678 & 116 & GIMAP7|168537 & 0.544\\ 
33 & ITGB2|3689 & 0.674 & 117 & GMFG|9535 & 0.544\\ 
34 & FCER1G|2207 & 0.674 & 118 & SRGN|5552 & 0.542\\ 
35 & DOK2|9046 & 0.672 & 119 & CCR1|1230 & 0.542\\ 
36 & SNX20|124460 & 0.672 & 120 & LYZ|4069 & 0.541\\ 
37 & C1QB|713 & 0.668 & 121 & TLR7|51284 & 0.541\\ 
38 & GIMAP4|55303 & 0.667 & 122 & IFI30|10437 & 0.54\\ 
39 & CD2|914 & 0.665 & 123 & SLAMF1|6504 & 0.538\\ 
40 & AIF1|199 & 0.665 & 124 & FCGR1B|2210 & 0.537\\ 
41 & IL21R|50615 & 0.656 & 125 & LPXN|9404 & 0.536\\ 
42 & TYROBP|7305 & 0.656 & 126 & PTPN22|26191 & 0.535\\ 
43 & CD48|962 & 0.652 & 127 & LY86|9450 & 0.534\\ 
44 & APBB1IP|54518 & 0.65 & 128 & HLA-DMB|3109 & 0.534\\ 
45 & C1QC|714 & 0.649 & 129 & KIAA0748|9840 & 0.533\\ 
46 & CD3E|916 & 0.644 & 130 & IL16|3603 & 0.532\\ 
47 & GIMAP5|55340 & 0.641 & 131 & CSF2RB|1439 & 0.532\\ 
48 & RASAL3|64926 & 0.64 & 132 & CD3D|915 & 0.532\\ 
49 & SPN|6693 & 0.637 & 133 & WIPF1|7456 & 0.531\\ 
50 & C3AR1|719 & 0.635 & 134 & SIGLEC7|27036 & 0.53\\ 
51 & GPR65|8477 & 0.631 & 135 & DOK3|79930 & 0.529\\ 
52 & FGL2|10875 & 0.629 & 136 & SIRPG|55423 & 0.527\\ 
53 & TAGAP|117289 & 0.629 & 137 & TIGIT|201633 & 0.525\\ 
54 & MNDA|4332 & 0.628 & 138 & RHOH|399 & 0.524\\ 
55 & EVI2A|2123 & 0.627 & 139 & ACAP1|9744 & 0.524\\ 
56 & CSF1R|1436 & 0.625 & 140 & CD247|919 & 0.523\\ 
57 & DOCK2|1794 & 0.623 & 141 & SLA2|84174 & 0.522\\ 
58 & IRF8|3394 & 0.621 & 142 & UBASH3A|53347 & 0.522\\ 
59 & SIGLEC10|89790 & 0.621 & 143 & NCF4|4689 & 0.52\\ 
60 & SAMSN1|64092 & 0.619 & 144 & GAB3|139716 & 0.52\\ 
61 & IKZF1|10320 & 0.618 & 145 & CD52|1043 & 0.519\\ 
62 & HLA-DPB1|3115 & 0.617 & 146 & CTSS|1520 & 0.518\\ 
63 & CD86|942 & 0.615 & 147 & ITGAX|3687 & 0.516\\ 
64 & SLAMF6|114836 & 0.614 & 148 & CCL5|6352 & 0.514\\ 
65 & TFEC|22797 & 0.611 & 149 & SIT1|27240 & 0.514\\ 
66 & CD84|8832 & 0.611 & 150 & PARVG|64098 & 0.514\\ 
67 & IGSF6|10261 & 0.608 & 151 & PYHIN1|149628 & 0.513\\ 
68 & SLCO2B1|11309 & 0.608 & 152 & NKG7|4818 & 0.512\\ 
69 & LILRB2|10288 & 0.608 & 153 & CD300A|11314 & 0.512\\ 
70 & HLA-DPA1|3113 & 0.606 & 154 & LOC100233209|100233209 & 0.512\\ 
71 & LAT2|7462 & 0.603 & 155 & GZMK|3003 & 0.511\\ 
72 & TNFAIP8L2|79626 & 0.601 & 156 & AOAH|313 & 0.509\\ 
73 & ARHGAP15|55843 & 0.596 & 157 & CD180|4064 & 0.509\\ 
74 & FAM78A|286336 & 0.594 & 158 & MS4A7|58475 & 0.508\\ 
75 & TLR8|51311 & 0.593 & 159 & GVIN1|387751 & 0.508\\ 
76 & ITK|3702 & 0.593 & 160 & CD33|945 & 0.507\\ 
77 & FCGR3A|2214 & 0.592 & 161 & FGD2|221472 & 0.506\\ 
78 & HCST|10870 & 0.591 & 162 & LY9|4063 & 0.505\\ 
79 & IL12RB1|3594 & 0.59 & 163 & MS4A4A|51338 & 0.505\\ 
80 & LRRC25|126364 & 0.588 & 164 & FMNL1|752 & 0.503\\ 
81 & PIK3R5|23533 & 0.584 & 165 & FGR|2268 & 0.502\\ 
82 & CXCR3|2833 & 0.584 & 166 & TRPV2|51393 & 0.502\\ 
83 & RNASE6|6039 & 0.583 & 167 & HCK|3055 & 0.501\\ 
84 & TNFRSF1B|7133 & 0.582 & 168 & WDFY4|57705 & 0.501\\ 
\hline
\multicolumn{6}{|l|}{\textbf{CIN mRNA attractor}}\\
\hline
1 & TPX2|22974 & 0.776 & 48 & SKA1|220134 & 0.604\\ 
2 & KIF4A|24137 & 0.753 & 49 & CENPF|1063 & 0.602\\ 
3 & KIFC1|3833 & 0.734 & 50 & NDC80|10403 & 0.598\\ 
4 & NCAPG|64151 & 0.732 & 51 & CEP55|55165 & 0.597\\ 
5 & BUB1|699 & 0.73 & 52 & CDC6|990 & 0.597\\ 
6 & NCAPH|23397 & 0.729 & 53 & BIRC5|332 & 0.596\\ 
7 & CDCA5|113130 & 0.725 & 54 & CDK1|983 & 0.59\\ 
8 & KIF2C|11004 & 0.725 & 55 & ARHGAP11A|9824 & 0.584\\ 
9 & PLK1|5347 & 0.723 & 56 & RAD54L|8438 & 0.583\\ 
10 & CENPA|1058 & 0.709 & 57 & STIL|6491 & 0.583\\ 
11 & TOP2A|7153 & 0.702 & 58 & CDC45|8318 & 0.581\\ 
12 & HJURP|55355 & 0.702 & 59 & DTL|51514 & 0.581\\ 
13 & BUB1B|701 & 0.698 & 60 & CDC25C|995 & 0.579\\ 
14 & KIF23|9493 & 0.691 & 61 & DEPDC1B|55789 & 0.569\\ 
15 & FOXM1|2305 & 0.69 & 62 & EPR1|8475 & 0.561\\ 
16 & MCM10|55388 & 0.687 & 63 & CCNB1|891 & 0.556\\ 
17 & KIF18B|146909 & 0.683 & 64 & ERCC6L|54821 & 0.55\\ 
18 & CCNA2|890 & 0.677 & 65 & MKI67|4288 & 0.55\\ 
19 & GTSE1|51512 & 0.676 & 66 & KIF18A|81930 & 0.548\\ 
20 & CKAP2L|150468 & 0.676 & 67 & SPC25|57405 & 0.546\\ 
21 & CCNB2|9133 & 0.676 & 68 & GSG2|83903 & 0.543\\ 
22 & DLGAP5|9787 & 0.674 & 69 & CDCA3|83461 & 0.543\\ 
23 & KIF11|3832 & 0.672 & 70 & CENPI|2491 & 0.541\\ 
24 & CDCA8|55143 & 0.666 & 71 & CENPE|1062 & 0.54\\ 
25 & KIF14|9928 & 0.663 & 72 & CDCA2|157313 & 0.537\\ 
26 & MELK|9833 & 0.662 & 73 & FANCI|55215 & 0.537\\ 
27 & NEK2|4751 & 0.651 & 74 & POLQ|10721 & 0.535\\ 
28 & AURKB|9212 & 0.649 & 75 & RAD51|5888 & 0.53\\ 
29 & PRC1|9055 & 0.646 & 76 & C15orf42|90381 & 0.526\\ 
30 & ASPM|259266 & 0.644 & 77 & KPNA2|3838 & 0.521\\ 
31 & KIF20A|10112 & 0.643 & 78 & ZWINT|11130 & 0.521\\ 
32 & EXO1|9156 & 0.642 & 79 & FAM72B|653820 & 0.519\\ 
33 & CDC20|991 & 0.642 & 80 & ESCO2|157570 & 0.516\\ 
34 & MYBL2|4605 & 0.64 & 81 & PLK4|10733 & 0.515\\ 
35 & RACGAP1|29127 & 0.633 & 82 & ASF1B|55723 & 0.514\\ 
36 & RRM2|6241 & 0.632 & 83 & ECT2|1894 & 0.513\\ 
37 & SGOL1|151648 & 0.631 & 84 & ESPL1|9700 & 0.513\\ 
38 & DEPDC1|55635 & 0.629 & 85 & LMNB1|4001 & 0.511\\ 
39 & ORC1L|4998 & 0.627 & 86 & SPAG5|10615 & 0.51\\ 
40 & TROAP|10024 & 0.625 & 87 & FAM64A|54478 & 0.51\\ 
41 & UBE2C|11065 & 0.62 & 88 & PTTG1|9232 & 0.508\\ 
42 & TTK|7272 & 0.62 & 89 & CASC5|57082 & 0.506\\ 
43 & SKA3|221150 & 0.614 & 90 & CDKN3|1033 & 0.505\\ 
44 & AURKA|6790 & 0.613 & 91 & UHRF1|29128 & 0.504\\ 
45 & KIF15|56992 & 0.612 & 92 & SHCBP1|79801 & 0.504\\ 
46 & NUSAP1|51203 & 0.609 & 93 & OIP5|11339 & 0.501\\ 
47 & NUF2|83540 & 0.608 & 94 & PBK|55872 & 0.501\\ 
\hline
\multicolumn{6}{|l|}{\textbf{MES mRNA attractor}}\\
\hline
1 & COL3A1|1281 & 0.798 & 31 & SFRP2|6423 & 0.572\\ 
2 & COL5A2|1290 & 0.775 & 32 & FNDC1|84624 & 0.567\\ 
3 & COL1A2|1278 & 0.771 & 33 & ISLR|3671 & 0.559\\ 
4 & THBS2|7058 & 0.753 & 34 & COL10A1|1300 & 0.554\\ 
5 & COL5A1|1289 & 0.746 & 35 & CRISPLD2|83716 & 0.551\\ 
6 & VCAN|1462 & 0.726 & 36 & COL8A1|1295 & 0.549\\ 
7 & COL6A3|1293 & 0.717 & 37 & BNC2|54796 & 0.545\\ 
8 & SPARC|6678 & 0.715 & 38 & LUM|4060 & 0.543\\ 
9 & AEBP1|165 & 0.709 & 39 & ANTXR1|84168 & 0.542\\ 
10 & FBN1|2200 & 0.705 & 40 & THY1|7070 & 0.539\\ 
11 & POSTN|10631 & 0.683 & 41 & ASPN|54829 & 0.537\\ 
12 & FAP|2191 & 0.654 & 42 & COL6A1|1291 & 0.537\\ 
13 & MMP2|4313 & 0.646 & 43 & NID2|22795 & 0.533\\ 
14 & COL1A1|1277 & 0.644 & 44 & COL11A1|1301 & 0.531\\ 
15 & PDGFRB|5159 & 0.641 & 45 & DACT1|51339 & 0.53\\ 
16 & LRRC15|131578 & 0.64 & 46 & FN1|2335 & 0.53\\ 
17 & ADAMTS2|9509 & 0.631 & 47 & LAMA4|3910 & 0.528\\ 
18 & ITGA11|22801 & 0.63 & 48 & SULF1|23213 & 0.526\\ 
19 & ADAM12|8038 & 0.628 & 49 & GPR124|25960 & 0.524\\ 
20 & OLFML2B|25903 & 0.622 & 50 & CCDC80|151887 & 0.524\\ 
21 & EMILIN1|11117 & 0.607 & 51 & MXRA5|25878 & 0.523\\ 
22 & COL6A2|1292 & 0.607 & 52 & OLFML1|283298 & 0.523\\ 
23 & TIMP2|7077 & 0.592 & 53 & CTHRC1|115908 & 0.52\\ 
24 & CDH11|1009 & 0.591 & 54 & PCOLCE|5118 & 0.516\\ 
25 & GLT8D2|83468 & 0.591 & 55 & ACTA2|59 & 0.51\\ 
26 & CTSK|1513 & 0.588 & 56 & GREM1|26585 & 0.509\\ 
27 & PRRX1|5396 & 0.587 & 57 & DCN|1634 & 0.508\\ 
28 & ADAMTS12|81792 & 0.585 & 58 & CALD1|800 & 0.503\\ 
29 & ANGPTL2|23452 & 0.584 & 59 & MSRB3|253827 & 0.501\\ 
30 & BGN|633 & 0.574 &  &  & \\ 
\hline
\multicolumn{6}{|l|}{\textbf{END mRNA attractor}}\\
\hline
1 & CDH5|1003 & 0.813 & 15 & RHOJ|57381 & 0.6\\ 
2 & ROBO4|54538 & 0.771 & 16 & BCL6B|255877 & 0.573\\ 
3 & CXorf36|79742 & 0.761 & 17 & TEK|7010 & 0.567\\ 
4 & CD34|947 & 0.733 & 18 & GPR116|221395 & 0.563\\ 
5 & CLEC14A|161198 & 0.688 & 19 & ACVRL1|94 & 0.56\\ 
6 & ARHGEF15|22899 & 0.673 & 20 & ECSCR|641700 & 0.551\\ 
7 & CD93|22918 & 0.672 & 21 & VWF|7450 & 0.549\\ 
8 & LDB2|9079 & 0.67 & 22 & KDR|3791 & 0.547\\ 
9 & ELTD1|64123 & 0.667 & 23 & EMCN|51705 & 0.533\\ 
10 & MYCT1|80177 & 0.661 & 24 & PTPRB|5787 & 0.517\\ 
11 & TIE1|7075 & 0.661 & 25 & NOTCH4|4855 & 0.515\\ 
12 & S1PR1|1901 & 0.653 & 26 & ERG|2078 & 0.514\\ 
13 & ESAM|90952 & 0.64 & 27 & PECAM1|5175 & 0.507\\ 
14 & PCDH12|51294 & 0.605 &  &  & \\ 
\hline
\multicolumn{6}{|l|}{\textbf{\textit{AHSA2} mRNA attractor}}\\
\hline
1 & AHSA2|130872 & 0.775 & 15 & CSAD|51380 & 0.545\\ 
2 & LOC91316|91316 & 0.641 & 16 & GOLGA8B|440270 & 0.539\\ 
3 & PILRB|29990 & 0.63 & 17 & GOLGA8A|23015 & 0.535\\ 
4 & ZNF767|79970 & 0.63 & 18 & NCRNA00105|80161 & 0.522\\ 
5 & TTLL3|26140 & 0.612 & 19 & CROCCL2|114819 & 0.521\\ 
6 & CCNL2|81669 & 0.606 & 20 & ?|155060 & 0.52\\ 
7 & PABPC1L|80336 & 0.606 & 21 & AGAP4|119016 & 0.52\\ 
8 & LENG8|114823 & 0.602 & 22 & LOC100272228|100272228 & 0.517\\ 
9 & CHKB-CPT1B|386593 & 0.595 & 23 & LUC7L3|51747 & 0.517\\ 
10 & SEC31B|25956 & 0.585 & 24 & C1orf104|284618 & 0.512\\ 
11 & NKTR|4820 & 0.57 & 25 & NCRNA00201|284702 & 0.51\\ 
12 & AGAP6|414189 & 0.569 & 26 & WASH7P|653635 & 0.504\\ 
13 & PDXDC2|283970 & 0.545 & 27 & LOC100131434|100131434 & 0.504\\ 
14 & HERC2P2|400322 & 0.545 &  &  & \\ 
\hline
\multicolumn{6}{|l|}{\textbf{IFIT mRNA attractor}}\\
\hline
1 & IFIT3|3437 & 0.779 & 11 & IFIT2|3433 & 0.594\\ 
2 & MX1|4599 & 0.761 & 12 & XAF1|54739 & 0.59\\ 
3 & OAS2|4939 & 0.757 & 13 & OASL|8638 & 0.589\\ 
4 & RSAD2|91543 & 0.753 & 14 & ISG15|9636 & 0.565\\ 
5 & CMPK2|129607 & 0.741 & 15 & HERC6|55008 & 0.56\\ 
6 & IFIT1|3434 & 0.732 & 16 & OAS3|4940 & 0.552\\ 
7 & IFI44L|10964 & 0.721 & 17 & IFIH1|64135 & 0.552\\ 
8 & IFI44|10561 & 0.695 & 18 & DDX58|23586 & 0.54\\ 
9 & IFI6|2537 & 0.644 & 19 & DDX60|55601 & 0.512\\ 
10 & OAS1|4938 & 0.613 &  &  & \\ 
\hline
\multicolumn{6}{|l|}{\textbf{\textit{WDR38} mRNA attractor}}\\
\hline
1 & WDR38|401551 & 0.677 & 11 & C9orf171|389799 & 0.563\\ 
2 & YSK4|80122 & 0.623 & 12 & CCDC135|84229 & 0.559\\ 
3 & ROPN1L|83853 & 0.613 & 13 & C1orf192|257177 & 0.557\\ 
4 & C1orf194|127003 & 0.603 & 14 & CAPSL|133690 & 0.554\\ 
5 & MORN5|254956 & 0.597 & 15 & ZBBX|79740 & 0.553\\ 
6 & WDR16|146845 & 0.587 & 16 & CCDC42B|387885 & 0.523\\ 
7 & RSPH4A|345895 & 0.577 & 17 & C1orf92|149499 & 0.522\\ 
8 & FAM183A|440585 & 0.574 & 18 & C2orf39|92749 & 0.511\\ 
9 & ZMYND10|51364 & 0.565 & 19 & DNAH12|201625 & 0.508\\ 
10 & DNAI1|27019 & 0.564 & 20 & RSPH1|89765 & 0.505\\ 
\hline
\multicolumn{6}{|l|}{\textbf{MHC Class II GL mRNA attractor -- chr6p21.32}}\\
\hline
1 & HLA-DPA1|3113 & 0.865 & 6 & HLA-DMB|3109 & 0.734\\ 
2 & HLA-DRA|3122 & 0.859 & 7 & HLA-DOA|3111 & 0.674\\ 
3 & HLA-DPB1|3115 & 0.787 & 8 & HLA-DQA1|3117 & 0.614\\ 
4 & HLA-DRB1|3123 & 0.76 & 9 & HLA-DRB5|3127 & 0.503\\ 
5 & HLA-DMA|3108 & 0.746 & 10 & HLA-DQB1|3119 & 0.387\\ 
\hline
\multicolumn{6}{|l|}{\textbf{GIMAP GL mRNA attractor -- chr7q36.1}}\\
\hline
1 & GIMAP4|55303 & 0.553 & 6 & GIMAP1|170575 & 0.471\\ 
2 & GIMAP7|168537 & 0.55 & 7 & GIMAP2|26157 & 0.269\\ 
3 & GIMAP6|474344 & 0.547 & 8 & TMEM176B|28959 & 0.179\\ 
4 & GIMAP5|55340 & 0.539 & 9 & TMEM176A|55365 & 0.143\\ 
5 & GIMAP8|155038 & 0.495 & 10 & ZNF777|27153 & 0.127\\ 
\hline
\multicolumn{6}{|l|}{\textbf{chr8q24.3 GL mRNA attractor}}\\
\hline
1 & SHARPIN|81858 & 0.585 & 6 & EXOSC4|54512 & 0.339\\ 
2 & HSF1|3297 & 0.575 & 7 & SCRIB|23513 & 0.307\\ 
3 & TIGD5|84948 & 0.535 & 8 & CYHR1|50626 & 0.294\\ 
4 & GPR172A|79581 & 0.482 & 9 & MAF1|84232 & 0.246\\ 
5 & ZC3H3|23144 & 0.42 & 10 & PUF60|22827 & 0.234\\ 

\end{longtable}

%
%

\setcounter{table}{0}
\makeatletter 
\renewcommand{\thetable}{S\@arabic\c@table} 
\makeatother

\setlength{\LTleft}{-0.9in}
\setlength{\LTright}{\LTleft}
\begin{longtable}{|l|c|c|c|l|c|c|c|}
\hline 
\textbf{Rank} & \textbf{Probe} & \textbf{GeneSymbol-Chr-Location} & \textbf{Score} & 
\textbf{Rank} & \textbf{Probe} & \textbf{GeneSymbol-Chr-Location} & \textbf{Score}\\ 
\hline 
\endfirsthead

\multicolumn{8}{c}%
{{\bfseries \tablename \thetable{} -- continued from previous page}} \\
\hline 
\textbf{Rank} & \textbf{Probe} & \textbf{GeneSymbol-Chr-Location} & \textbf{Score} & 
\textbf{Rank} & \textbf{Probe} & \textbf{GeneSymbol-Chr-Location} & \textbf{Score}\\ 
\hline 
\endhead

\hline \multicolumn{8}{r}{{Continued on next page...}} \\
\endfoot

\hline \hline
\endlastfoot

\hline
\multicolumn{8}{|l|}{\textbf{RMND1 methylation attractor}}\\
\hline
1 & cg00510787 & RMND1-6-151814639 & 0.842 & 61 & cg18325289 & USP48-1-21982534 & 0.586\\ 
2 & cg01087382 & MAP3K7-6-91353911 & 0.836 & 62 & cg17607973 & MEPCE-7-99865344 & 0.586\\ 
3 & cg08965527 & DNAAF1-16-82735714 & 0.822 & 63 & cg11628034 & RPS16-19-44618744 & 0.579\\ 
4 & cg08793459 & PTRH2-17-55139429 & 0.819 & 64 & cg25409040 & CSTF3-11-33139919 & 0.579\\ 
5 & cg14037413 & ZNF143-11-9439170 & 0.817 & 65 & cg09747578 & LRRC41-1-46541663 & 0.576\\ 
6 & cg03627896 & NA-NA-30841835 & 0.805 & 66 & cg16016641 & C20orf4-20-34287611 & 0.574\\ 
7 & cg03169527 & TAMM41-3-11863582 & 0.804 & 67 & cg26979012 & PSMC6-14-52243814 & 0.574\\ 
8 & cg11368578 & CDK5-7-150385869 & 0.787 & 68 & cg24356797 & UBA52-19-18543387 & 0.573\\ 
9 & cg19233923 & OTUB1-11-63510174 & 0.781 & 69 & cg26258330 & ZFAND1-8-82795685 & 0.572\\ 
10 & cg25742201 & AATF-17-32380976 & 0.78 & 70 & cg22468803 & ANKRD12-18-9126381 & 0.571\\ 
11 & cg20684973 & TAF5-10-105117622 & 0.779 & 71 & cg01439983 & PEX13-2-61097497 & 0.567\\ 
12 & cg06719391 & NA-NA-2510906 & 0.776 & 72 & cg02776251 & USP15-12-60940586 & 0.565\\ 
13 & cg23179321 & RPP38-10-15178799 & 0.775 & 73 & cg19469297 & TBRG4-7-45117965 & 0.562\\ 
14 & cg25346576 & NSRP1-17-25467978 & 0.756 & 74 & cg26477793 & ACTR1A-10-104252075 & 0.561\\ 
15 & cg04305134 & ANXA7-10-74843380 & 0.743 & 75 & cg27196102 & COG7-16-23372214 & 0.56\\ 
16 & cg24568646 & CCT8-21-29368109 & 0.739 & 76 & cg01998146 & PSIP1-9-15501358 & 0.558\\ 
17 & cg23347958 & DHX8-17-38916826 & 0.731 & 77 & cg24695828 & ZNF566-19-41672097 & 0.557\\ 
18 & cg08209724 & ZNF207-17-27701251 & 0.729 & 78 & cg12790134 & VCPIP1-8-67741341 & 0.553\\ 
19 & cg15749322 & ANKRD42-11-82582847 & 0.728 & 79 & cg16141690 & C14orf119-14-22633555 & 0.551\\ 
20 & cg26538116 & LRPPRC-2-44076832 & 0.726 & 80 & cg01466020 & YY1AP1-1-153925818 & 0.55\\ 
21 & cg26446827 & ZNF133-20-18216978 & 0.726 & 81 & cg15875120 & FAM188A-10-15942611 & 0.547\\ 
22 & cg09047884 & TTLL1-22-41814928 & 0.725 & 82 & cg11084020 & TAX1BP1-7-27746092 & 0.544\\ 
23 & cg25418748 & RUFY1-5-178909842 & 0.724 & 83 & cg27546682 & STK40-1-36624447 & 0.544\\ 
24 & cg24654547 & DUS2L-16-66614666 & 0.72 & 84 & cg15009698 & AP3B1-5-77626250 & 0.542\\ 
25 & cg12662162 & C6orf106-6-34772865 & 0.711 & 85 & cg03100196 & ZYX-7-142787883 & 0.542\\ 
26 & cg17718515 & IPO9-1-200065440 & 0.698 & 86 & cg19747852 & TRA2A-7-23538218 & 0.541\\ 
27 & cg21605986 & KIAA1191-5-175721331 & 0.691 & 87 & cg13022174 & SS18-18-21925057 & 0.54\\ 
28 & cg06611744 & RASL11A-13-26742618 & 0.688 & 88 & cg17159242 & RPS7-2-3600935 & 0.539\\ 
29 & cg08871016 & HSF2BP-21-43903660 & 0.682 & 89 & cg05856931 & MSI2-17-52688096 & 0.536\\ 
30 & cg11011602 & RPL7A-9-135204850 & 0.68 & 90 & cg08529259 & GCSH-16-79686870 & 0.536\\ 
31 & cg07979357 & IL27RA-19-14003353 & 0.68 & 91 & cg16769442 & MAGI3-1-113735582 & 0.536\\ 
32 & cg25658980 & PIH1D1-19-54646816 & 0.679 & 92 & cg24478630 & MOGS-2-74546204 & 0.535\\ 
33 & cg05577173 & PAPD4-5-78944434 & 0.675 & 93 & cg16425577 & DENND3-8-142208836 & 0.532\\ 
34 & cg03843852 & PIGB-15-53398391 & 0.672 & 94 & cg07209631 & PPM1K-4-89424752 & 0.532\\ 
35 & cg23213688 & EXOC8-1-229540414 & 0.667 & 95 & cg27227786 & DGKE-17-52266264 & 0.53\\ 
36 & cg21643860 & TUT1-11-62115390 & 0.663 & 96 & cg08158331 & SH2D6-2-85499059 & 0.53\\ 
37 & cg16854524 & LIN54-4-84150926 & 0.658 & 97 & cg18142353 & LRP6-12-12311841 & 0.528\\ 
38 & cg20188282 & GTF3C5-9-134896021 & 0.656 & 98 & cg08613513 & ORC1-1-52642648 & 0.527\\ 
39 & cg21589280 & DDAH1-1-85702739 & 0.654 & 99 & cg12835684 & MFSD5-12-51931914 & 0.526\\ 
40 & cg20908993 & ATP6V1D-14-66896131 & 0.65 & 100 & cg08717396 & HIST1H2AG-6-27208754 & 0.524\\ 
41 & cg14191109 & SLC25A16-10-69957444 & 0.649 & 101 & cg19857457 & RPL17-18-45272945 & 0.522\\ 
42 & cg27072323 & CAP1-1-40279445 & 0.647 & 102 & cg17868994 & NIPSNAP3B-9-106566104 & 0.521\\ 
43 & cg21666675 & CHCHD4-3-14140844 & 0.646 & 103 & cg15806518 & PSMB7-9-126217638 & 0.521\\ 
44 & cg08358671 & IVNS1ABP-1-183553430 & 0.64 & 104 & cg03863549 & NOL6-9-33464004 & 0.521\\ 
45 & cg23749163 & KIAA1737-14-76633721 & 0.639 & 105 & cg01449415 & ZNF213-16-3124812 & 0.52\\ 
46 & cg21452766 & CTH-1-70649676 & 0.636 & 106 & cg11371394 & TGFBRAP1-2-105313015 & 0.519\\ 
47 & cg03532005 & PSPH-7-56086522 & 0.635 & 107 & cg08554462 & NA-NA-59137891 & 0.519\\ 
48 & cg07031532 & OAZ2-15-62782737 & 0.632 & 108 & cg09378940 & LACTB2-8-71743860 & 0.518\\ 
49 & cg09507928 & IK-5-140007668 & 0.63 & 109 & cg25264554 & XRN1-3-143649479 & 0.518\\ 
50 & cg05797656 & DDX42-17-59205084 & 0.629 & 110 & cg10095719 & NRP1-10-33663793 & 0.517\\ 
51 & cg23526055 & TCEB1-8-75047348 & 0.627 & 111 & cg22721186 & ATAD2-8-124478099 & 0.517\\ 
52 & cg08376864 & HRASLS-3-194441420 & 0.622 & 112 & cg16268429 & NADKD1-5-36277893 & 0.516\\ 
53 & cg06422693 & OCIAD1-4-48528344 & 0.618 & 113 & cg21546057 & FIBP-11-65412902 & 0.516\\ 
54 & cg20760063 & NA-NA-38531106 & 0.616 & 114 & cg21824902 & WDTC1-1-27433419 & 0.512\\ 
55 & cg14817848 & DUSP16-12-12606825 & 0.615 & 115 & cg22176017 & POLR2G-11-62286036 & 0.511\\ 
56 & cg23653187 & PNPLA3-22-42650590 & 0.603 & 116 & cg14711201 & SKP2-5-36188193 & 0.511\\ 
57 & cg04883450 & DIMT1-5-61735714 & 0.601 & 117 & cg07820214 & RUVBL1-3-129325679 & 0.511\\ 
58 & cg22464182 & ACVR2A-2-148319095 & 0.595 & 118 & cg15452426 & DFFA-1-10454817 & 0.507\\ 
59 & cg21097640 & POLR1D-13-27093548 & 0.593 & 119 & cg16050957 & RBBP6-16-24458244 & 0.506\\ 
60 & cg01190915 & MT2A-16-55200262 & 0.587 & 120 & cg00115714 & AMD1-6-111302805 & 0.503\\ 
\hline
\multicolumn{8}{|l|}{\textbf{M+ methylation attractor}}\\
\hline
1 & cg13928306 & NA-NA-3102398 & 0.72 & 41 & cg27238470 & EDN2-1-41722912 & 0.566\\ 
2 & cg24620905 & MTMR11-1-148175405 & 0.719 & 42 & cg03684977 & GRB7-17-35147329 & 0.565\\ 
3 & cg27324619 & NA-NA-29848860 & 0.713 & 43 & cg09243900 & RAB25-1-154297468 & 0.562\\ 
4 & cg12603560 & TNKS1BP1-11-56846646 & 0.712 & 44 & cg26531804 & SPINT1-15-38923139 & 0.559\\ 
5 & cg08775230 & C11orf52-11-111294703 & 0.709 & 45 & cg08463485 & ILDR1-3-123223617 & 0.556\\ 
6 & cg07705835 & IL17RC-3-9934128 & 0.703 & 46 & cg10917602 & HSD3B7-16-30904131 & 0.553\\ 
7 & cg24765079 & NA-NA-67329931 & 0.694 & 47 & cg22585988 & PVRL4-1-159325951 & 0.552\\ 
8 & cg19258882 & ERBB3-12-54759072 & 0.683 & 48 & cg01919208 & LAMB2-3-49145500 & 0.551\\ 
9 & cg09152089 & IL22RA1-1-24342151 & 0.68 & 49 & cg10212621 & HMGCS2-1-120113172 & 0.551\\ 
10 & cg05697249 & C11orf52-11-111294903 & 0.677 & 50 & cg17186163 & C10orf10-10-44794323 & 0.548\\ 
11 & cg18053607 & NA-NA-29848963 & 0.663 & 51 & cg18988110 & PRR15L-17-43390342 & 0.547\\ 
12 & cg14036856 & C1orf210-1-43524150 & 0.65 & 52 & cg23349242 & SUSD2-22-22907448 & 0.545\\ 
13 & cg04245402 & C19orf21-19-702241 & 0.649 & 53 & cg05225996 & NA-NA-3103018 & 0.544\\ 
14 & cg20324165 & KRT8-12-51585412 & 0.638 & 54 & cg13530039 & CHRM1-11-62446133 & 0.542\\ 
15 & cg13439730 & PRSS8-16-31054500 & 0.63 & 55 & cg20484352 & FAM114A1-4-38546174 & 0.537\\ 
16 & cg24433189 & SSTR5-16-1068690 & 0.622 & 56 & cg09440340 & MAB21L3-1-116455938 & 0.535\\ 
17 & cg16176600 & FRK-6-116488302 & 0.617 & 57 & cg22764925 & GGT1-22-23309964 & 0.533\\ 
18 & cg25370441 & ARHGEF38-4-106692682 & 0.612 & 58 & cg14528319 & GIPC1-19-14468713 & 0.532\\ 
19 & cg09307264 & INCA1-17-4843005 & 0.61 & 59 & cg07947930 & PRELP-1-201711568 & 0.532\\ 
20 & cg17826679 & SLC44A2-19-10597038 & 0.609 & 60 & cg22780475 & CBLC-19-49973366 & 0.53\\ 
21 & cg00698688 & SULT2B1-19-53747244 & 0.603 & 61 & cg05245515 & SLC39A2-14-20537113 & 0.526\\ 
22 & cg09548084 & SLC35B3-6-8381217 & 0.602 & 62 & cg17740645 & GRB7-17-35147939 & 0.524\\ 
23 & cg21663431 & SLC44A2-19-10597355 & 0.601 & 63 & cg18565355 & ESRP1-8-95723050 & 0.523\\ 
24 & cg25946374 & IL22RA1-1-24342378 & 0.599 & 64 & cg25415932 & OGG1-3-9766051 & 0.521\\ 
25 & cg22580512 & NCOR2-12-123568427 & 0.59 & 65 & cg02537838 & C20orf151-20-60435990 & 0.52\\ 
26 & cg16986846 & SCGB2A1-11-61732750 & 0.59 & 66 & cg19923326 & NA-NA-2785354 & 0.519\\ 
27 & cg05517572 & STAP2-19-4289769 & 0.588 & 67 & cg23165541 & DAPK2-15-62125217 & 0.517\\ 
28 & cg16787352 & ANKRD9-14-102045158 & 0.587 & 68 & cg03599338 & SUSD2-22-22907315 & 0.516\\ 
29 & cg24210717 & GPD1-12-48784094 & 0.584 & 69 & cg01835489 & KRT8-12-51585577 & 0.513\\ 
30 & cg27126442 & ARHGEF38-4-106693255 & 0.583 & 70 & cg03003745 & CXCL17-19-47640007 & 0.513\\ 
31 & cg19759064 & PHKG1-7-56128181 & 0.579 & 71 & cg00451635 & EMP2-16-10582531 & 0.512\\ 
32 & cg05228408 & CLCN6-1-11787939 & 0.579 & 72 & cg23323671 & STMN1-1-26106210 & 0.508\\ 
33 & cg14322224 & DDAH1-1-85703814 & 0.578 & 73 & cg04806409 & TFF3-21-42608574 & 0.508\\ 
34 & cg12894126 & UCK1-9-133395746 & 0.577 & 74 & cg03109701 & RNFT2-12-115659141 & 0.508\\ 
35 & cg09871043 & PKHD1-6-52060320 & 0.575 & 75 & cg22820108 & NCOR2-12-123569171 & 0.508\\ 
36 & cg02293044 & GAS2L1-22-28033543 & 0.574 & 76 & cg08445039 & FKBP9-7-32964184 & 0.507\\ 
37 & cg02237119 & WBSCR27-7-72894350 & 0.574 & 77 & cg18632631 & TNK1-17-7224773 & 0.504\\ 
38 & cg00480115 & FXYD3-19-40298717 & 0.57 & 78 & cg01119135 & C1orf116-1-205272148 & 0.503\\ 
39 & cg24751129 & GNMT-6-43036898 & 0.569 & 79 & cg21201572 & AGR2-7-16811131 & 0.501\\ 
40 & cg04001668 & GPR56-16-56211848 & 0.566 & 80 & cg24835159 & RNF43-17-53849553 & 0.501\\ 
\hline
\multicolumn{8}{|l|}{\textbf{M- methylation attractor}}\\
\hline
1 & cg10590292 & BIN2-12-50003941 & 0.72 & 15 & cg15352829 & PLD4-14-104462063 & 0.576\\ 
2 & cg20792833 & PTPRCAP-11-66961771 & 0.659 & 16 & cg07380416 & CD6-11-60495748 & 0.576\\ 
3 & cg23612220 & TNFAIP8L2-1-149395922 & 0.653 & 17 & cg23953831 & CD101-1-117345939 & 0.567\\ 
4 & cg16927606 & IGFLR1-19-40925164 & 0.64 & 18 & cg03569637 & FAM113B-12-45895624 & 0.559\\ 
5 & cg05596756 & FAM113B-12-45896487 & 0.635 & 19 & cg14145194 & ICAM3-19-10311022 & 0.557\\ 
6 & cg09902130 & CD6-11-60495754 & 0.61 & 20 & cg26683005 & CARD8-19-53444802 & 0.553\\ 
7 & cg20425130 & KLHL6-3-184755939 & 0.602 & 21 & cg07732037 & MPHOSPH9-12-122273780 & 0.55\\ 
8 & cg18384097 & PTPN7-1-200396189 & 0.601 & 22 & cg11600161 & TBC1D10C-11-66928161 & 0.535\\ 
9 & cg17936488 & FAM78A-9-133141340 & 0.601 & 23 & cg01623438 & CTSZ-20-57016289 & 0.529\\ 
10 & cg25671438 & ACAP1-17-7180947 & 0.598 & 24 & cg23093496 & C16orf54-16-29664824 & 0.523\\ 
11 & cg16609957 & SHROOM1-5-132189766 & 0.593 & 25 & cg27377213 & PPP1R16B-20-36867217 & 0.519\\ 
12 & cg14519350 & OSM-22-28991949 & 0.588 & 26 & cg15691199 & CEBPE-14-22659259 & 0.503\\ 
13 & cg16509569 & NCKAP1L-12-53177901 & 0.582 & 27 & cg15551881 & TRAF1-9-122728536 & 0.503\\ 
14 & cg07973967 & CD79B-17-59363339 & 0.578 &  &  &  & \\ 

\end{longtable}

%
%

\setcounter{table}{0}
\makeatletter 
\renewcommand{\thetable}{S\@arabic\c@table} 
\makeatother

\setlength{\LTleft}{0in}
\setlength{\LTright}{\LTleft}
\begin{longtable}{|l|c|c|l|c|c|}

\hline 
\textbf{Rank} & \textbf{miRNA family} & \textbf{Score} & \textbf{Rank} & \textbf{miRNA family} & \textbf{Score}\\ 
\hline 
\endfirsthead

\multicolumn{6}{c}%
{{\bfseries \tablename \thetable{} -- continued from previous page}} \\
\hline 
\textbf{Rank} & \textbf{miRNA family} & \textbf{Score} & \textbf{Rank} & \textbf{miRNA family} & \textbf{Score}\\ 
\hline 
\endhead

\hline \multicolumn{6}{r}{{Continued on next page...}} \\
\endfoot

\hline \hline
\endlastfoot

\hline
\multicolumn{6}{|l|}{\textbf{mir-127 miRNA attractor}}\\
\hline
1 & hsa-mir-127 & 0.854 & 9 & hsa-mir-654 & 0.615\\ 
2 & hsa-mir-134 & 0.835 & 10 & hsa-mir-431 & 0.574\\ 
3 & hsa-mir-379 & 0.803 & 11 & hsa-mir-493 & 0.561\\ 
4 & hsa-mir-409 & 0.742 & 12 & hsa-mir-337 & 0.554\\ 
5 & hsa-mir-382 & 0.686 & 13 & hsa-mir-487 & 0.528\\ 
6 & hsa-mir-758 & 0.678 & 14 & hsa-mir-889 & 0.516\\ 
7 & hsa-mir-381 & 0.649 & 15 & hsa-mir-154 & 0.512\\ 
8 & hsa-mir-370 & 0.616 &  &  & \\ 
\hline
\multicolumn{6}{|l|}{\textbf{mir-509 miRNA attractor}}\\
\hline
1 & hsa-mir-509 & 0.899 & 3 & hsa-mir-508 & 0.81\\ 
2 & hsa-mir-514 & 0.824 &  &  & \\ 
\hline
\multicolumn{6}{|l|}{\textbf{mir-144 miRNA attractor}}\\
\hline
1 & hsa-mir-144 & 0.89 & 3 & hsa-mir-486 & 0.678\\ 
2 & hsa-mir-451 & 0.877 &  &  & \\ 
\end{longtable}

%
%

\setcounter{table}{0}
\makeatletter 
\renewcommand{\thetable}{S\@arabic\c@table} 
\makeatother

\begin{longtable}{|l|c|c|l|c|c|}

\hline 
\textbf{Rank} & \textbf{Protein} & \textbf{Score} & \textbf{Rank} & \textbf{Protein} & \textbf{Score}\\ 
\hline 
\endfirsthead

\multicolumn{6}{c}%
{{\bfseries \tablename \thetable{} -- continued from previous page}} \\
\hline 
\textbf{Rank} & \textbf{Protein} & \textbf{Score} & \textbf{Rank} & \textbf{Protein} & \textbf{Score}\\ 
\hline 
\endhead

\hline \multicolumn{6}{r}{{Continued on next page...}} \\
\endfoot

\hline \hline
\endlastfoot

\hline
\multicolumn{6}{|l|}{\textbf{c-Met protein attractor}}\\
\hline
1 & c-Met-M-C & 0.877 & 4 & Caspase-8-M-C & 0.741\\ 
2 & Snail-M-C & 0.829 & 5 & ERCC1-M-C & 0.674\\ 
3 & PARP\_cleaved-M-C & 0.763 & 6 & Rb-M-V & 0.567\\ 
\hline
\multicolumn{6}{|l|}{\textbf{Akt protein attractor}}\\
\hline
1 & Akt-R-V & 0.6 & 3 & STAT5-alpha-R-V & 0.519\\ 
2 & Tuberin-R-C & 0.552 &  &  & \\ 
\label{tab:tabS1}
\end{longtable}

\end{supptable}

\end{raggedright}
\end{document}